\documentclass[preprint2]{proto}
\usepackage{times}
\newcommand{\refs}{\par\noindent\hangindent=1pc\hangafter=1}
\def\solmas{{M$_\odot$}}
\def\simless{\mathbin{\lower 3pt\hbox
   {$\rlap{\raise 5pt\hbox{$\char'074$}}\mathchar"7218$}}}   
\def\simgreat{\mathbin{\lower 3pt\hbox
   {$\rlap{\raise 5pt\hbox{$\char'076$}}\mathchar"7218$}}}   
\def\etal{{\rm et al.}}
\def\etal{{et al.}}
\def\mj {M_{\rm Jeans}}
\def\rj {R_{\rm Jeans}}

\def\solmas{{M$_\odot$}}
\def\solm{{M_\odot}}

\def\vrel {v_{\rm rel}}
\def\tff {t_{\rm ff}}

\def\ms {M_*}
\def\msi {M_0}

\def\Rtidal {R_{\rm tidal}}
\def\racc {R_{\rm acc}}

\def\macc {\dot M_*}
\def\menc {M_{\rm enc}}
\def\Rclus {R_*}
\def\Rbh {R_{\rm BH}}

\def\araa {{\sl Ann. Rev. Astron. Astrophys.}}
\begin{document}

\title{\textbf{\LARGE The Origin of the Initial Mass Function}}

\author {\textbf{\large Ian A. Bonnell}}
\affil{\small\em University of St Andrews}
\author {\textbf{\large Richard B. Larson}}
\affil{\small\em Yale University}
\author {\textbf{\large Hans Zinnecker}}
\affil{\small\em Astrophysikalisches Institut Potsdam}

\begin{abstract}
\baselineskip = 11pt
\leftskip = 0.65in 
\rightskip = 0.65in
\parindent=1pc
{\small 
We review recent advances in our understanding of the origin of the initial 
mass function (IMF). We emphasize the
use of numerical simulations to investigate  how each physical process
involved in star formation affects the resulting IMF. We stress that it is insufficient 
to just reproduce the IMF, but that any 
successful model needs to account for the many observed properties of star forming regions 
including clustering, mass segregation and binarity. 
Fragmentation involving
the interplay of gravity, turbulence, and thermal effects
is probably responsible for setting the characteristic stellar mass. 
Low-mass stars and brown dwarfs can form through the fragmentation of dense filaments
and disks,
possibly followed by early ejection from these dense environments which truncates
their growth in mass.
Higher-mass stars and the Salpeter-like slope of the 
IMF are most
likely formed through continued accretion in a clustered environment. 
The effects of feedback and magnetic fields on the origin of the IMF are still largely unclear. 
Lastly, we discuss a 
number of outstanding problems that need to be addressed in order to develop a complete 
theory for the origin of the IMF.
\\~\\~\\~}

\end{abstract}

\section{\textbf{INTRODUCTION}}

One of the main goals for a theory of star formation is to understand the origin
of the stellar initial mass function (IMF). There has been considerable observational
work establishing the general form of the IMF (e.g., {\sl Scalo,} 1986, 1998; {\sl Kroupa,}~2001,~2002; 
{\sl Reid \etal,}~2002; {\sl Chabrier,} 2003), but as yet we do not have a
clear understanding of the physics that determines the distribution of stellar masses.
The aim of this chapter is to review the physical processes that are most likely
involved and to discuss observational tests that can be used to distinguish between them.

Understanding the origin of the IMF is crucial as
it includes the basic physics that determines our observable universe, the generation of the
chemical elements, the kinematic feedback into the ISM and overall the formation
and evolution of galaxies. Once we understand the origin of the IMF, we can also
contemplate how and when the IMF is likely to vary in certain environments such as the early
universe and the Galactic centre.

There have been many theoretical ideas  advanced to explain the IMF 
(cf. {\sl Miller and Scalo,}~1979; {\sl Silk and Takahashi,}~1979; {\sl Fleck,}~1982, {\sl Zinnecker,}~1982,1984; 
{\sl Elmegreen and Mathieu,}~1983;
{\sl Yoshii and Saio,}~1985; {\sl Silk,}~1995, {\sl Adams and 
Fatuzzo,}~1996, {\sl Elmegreen,}~1997, {\sl Clarke,}~1998; {\sl Meyer \etal,} 2000; {\sl Larson,}~2003,2005; 
{\sl Zinnecker \etal,}~1993; {\sl Zinnecker,} 2005; {\sl Corbelli \etal,}~ 2005; and references therein).
Most theories are 'successful'  in that they are able to derive a Salpeter-slope IMF 
({\sl Salpeter,}~1955) but generally  
 they have lacked significant predictive powers. The main problem is that it is far
too easy to develop a theory, typically involving many variables, that has as a goal
to explain a population distribution dependent on only one variable, the stellar mass.
There have  been a large number of analytical theories made to explain the IMF
and therefore the probability of any one of them being correct is relatively small.
It is thus imperative not only for a model to 'explain' the IMF, but also to develop 
secondary indicators  that can be used to assess its likelihood of contributing to a full theory.

Recent increases of computational power have implied that numerical simulations can now
include many of the relevant physical processes and be used to produce a measurable
IMF that can be compared with observations. This means that we no longer have to rely
on analytical arguments as to what individual processes can do but we can include
these processes into numerical simulations and can test what their effect is on 
star formation and the generation of an IMF.  
Most importantly, numerical simulations provide a wealth
of secondary information other than just an IMF and these can be taken to compare
directly with observed properties of young stars and star forming regions. We thus
concentrate in this  review on the use of numerical simulations to assess the importance
of the physical processes and guide us in our aim of developing a theory for the origin
of the initial mass function.

The initial mass function is generally categorized by a segmented power-law 
or a log-normal type mass distribution ({\sl Kroupa,} 2001; {\sl Chabrier,} 2003). 
For the sake of simplicity, we adopt the power-law formalism of the type
\begin{equation}
dN \propto m^{-\alpha} dm,
\end{equation}
but this should not be taken to mean that the IMF needs to be described in such a manner.
For clarity, it should be noted that IMFs are also commonly described in terms of a distribution
in log mass:
\begin{equation}
dN \propto m^{\Gamma} d({\rm log\ } m),
\end{equation}
where $\Gamma = -(\alpha - 1)$ ({\sl Scalo,} 1986).  
The {\sl Salpeter} (1955) slope for high-mass stars (see \S 2) is then $\alpha = 2.35$
or $\Gamma=-1.35$
We also note here that the critical values of $\alpha=2$, $\Gamma=-1$
occur when equal mass is present in each mass decade (for example 1 to 10 \solmas\
and 10 to 100\solmas).

\vspace{0.1cm} 

\section{\textbf{OBSERVED FEATURES}}
\vspace{0.1cm}

The most important feature of the IMF that we need to understand is the fact that
there is a characteristic mass for stars at slightly less than $ 1 \solm$. This is indicated
by the occurrence of a marked flattening of the IMF below one solar mass, such that the 
total mass does not diverge at either high or low stellar masses.
If we can explain this one basic feature, then we will have 
the foundation for a complete theory of star formation.
In terms of understanding the role of star formation in affecting the evolution of galaxies 
and their interstellar media,
it is the upper-mass Salpeter-like slope which is most important. The relative numbers
of massive stars determines the chemical and kinematic feedback of star formation. 
Other basic features of the IMF are most likely a lower, and potentially an upper
mass cut-off. 

\begin{figure}[h]
\epsscale{0.7}
\plotone{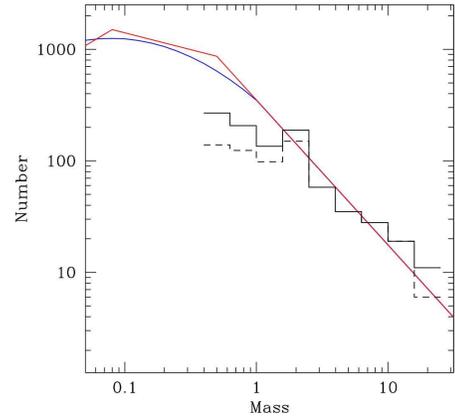}
\caption{\label{obsIMF} \small The IMF for NGC3603  ({\sl Stolte \etal},~2006)  is shown as a histogram in log mass ($dN({\rm log\ mass})$) for the completeness corrected (solid) and uncorrected (dashed) populations. For comparison, the {\sl Kroupa} (2001) segmented power-law and the {\sl Chabrier} (2003)  log-normal plus power-law IMFs
are also plotted in terms of log mass. 
}  
\end{figure}

One of the most remarkable features of IMF research is that the upper-mass Salpeter slope
has survived 50 years without significant revision (e.g., {\sl Salpeter,} 1955; 
{\sl Corbelli \etal},~2005).
At the same time, much work and debate has concentrated on understanding the low-mass
IMF (e.g., {\sl Reid \etal},~2002; {\sl Corbelli \etal},~2005 and references therein). 
It appears that the form of the IMF has converged to a certain degree
and is generally described as either a log-normal distribution with a
power-law tail or as a series of power-laws (see Figure~\ref{obsIMF}). For ease of description, the
IMF is generally given in the latter form, such as  the {\sl Kroupa} (2001) IMF 
\begin{equation} 
{\begin{array}{ll}
dN \propto m^{-2.3} dm  & (m \ge 0.5 \solm) \\
dN \propto m^{-1.3} dm  & (0.08 \le m \le 0.5 \solm) \\
dN \propto m^{-0.3} dm  & ( m \le 0.08 \solm) .
\end{array}}
\end{equation}

Observational studies of the IMF in regions of star formation (e.g., {\sl Meyer \etal},~2000,
{\sl Zinnecker \etal},~1993) have shown
that the IMF is set early in the star formation process. With the caveat that  stellar masses
(and ages) are difficult to extract during the pre-main sequence contraction phase, 
most young stellar regions have mass functions that follow a 'normal' IMF.  
In this case, the IMF
appears to be a (near) universal function of  star formation in our Galaxy.

One of the most pressing questions concerning the origin of the IMF is how early 
the mass distribution is set. Does this occur at the molecular cloud fragmentation stage
or does it occur afterwards due to gas accretion, feedback etc. ?
The possibility that the IMF is set at the pre-stellar core stage has received a 
significant boost from
observations of the clump-mass distributions in $\rho$ Oph, Serpens, Orion 
and others that appear 
to closely follow the stellar IMF
({\sl Motte \etal},~1998,~2001; {\sl Testi \etal},~1998; {\sl Johnstone \etal},~2000;
 see the chapter by  {\sl Lada \etal}). 
The main assumption is that there is a direct mapping of core to stellar masses. This
is uncertain for a number of reasons including the possibility that some or most of the cores
are gravitationally unbound ({\sl Johnstone \etal},~2000) and therefore may never form any stars.
If the cores do collapse, they are likely
to form binary or multiple stellar systems (e.g., {\sl Goodwin and Kroupa}, 2005) which would 
affect the resulting stellar IMF, at least for core masses $\ge 1 \solm$ ({\sl Lada}, 2006). Lastly,
in order for the clump-mass spectrum to match onto the IMF, none of the
extended mass in the system can become involved in the star formation process. 
{\sl Johnstone \etal},~(2004) report that in $\rho$ Oph only a few percent of the total mass is in the clumps.
The remaining mass also explains why the clump masses vary from study to study
as the masses are likely to depend on the exact location of the clump boundaries.

Another important question concerns the universality of the IMF, especially the
relative abundances of high and low-mass stars ({\sl Scalo}, 2005; {\sl Elmegreen}, 2004).
Although there have been occasional claims of top-heavy or truncated IMFs, they have
generally relied on unresolved stellar populations and have gone away when individual stars
are detected and counted.  At present, there are two cases which appear
to be more robust and worthy of consideration. They are both located near the Galactic
centre which may be an indication of the different physics there ({\sl Larson}, 2005). 
Firstly, there is the Arches cluster which appears to have a top-heavy IMF in the 
resolved population ({\sl Stolte \etal,}~2005).
Caveats are that this may be influenced by mass segregation in the cluster, incompleteness,
and perhaps unresolved binaries. The second case is the Galactic centre where the
massive stars are resolved ({\sl Paumard \etal},~2006), but there appears little evidence for 
a low-mass pre-main sequence
population based on expected X-ray fluxes and on dynamical mass estimates
({\sl Nayashin and Sunyaev}, 2005).

\section{RELEVANT OBSERVATIONAL CONSTRAINTS}

It is apparent that many models have been advanced to explain the origin of the IMF.
It is equally apparent that just being capable of reproducing the observed IMF is not
a sufficient condition. We need observational tests and secondary indicators
that can be used to distinguish between the models, be they current or in the future.
In theory, most if not all observed properties of young stars (discs, velocities, clusterings)
and star forming regions (mass distributions, kinematics) should
be explained by a complete model for the IMF. In practice, it is presently unclear
what the implications of many of the observed properties are.  Here we outline
a selection of potential tests that either can be used presently or are likely
to be usable in the next several years.

\subsection{Young stellar clusters}

It is becoming increasingly apparent that most stars form in groups and clusters
with the higher-mass stars forming almost exclusively in dense stellar environments.
Thus, models for the IMF need to account for the clustered nature of star formation  and
that the environment is likely to play an important role in determining the
stellar masses. For example, models for the IMF need to be able to reproduce
the cluster properties in terms of stellar densities, and spatial distributions of
lower and higher-mass stars.

One  question is whether there is a physical correlation between the star
forming environment and the formation of massive stars. 
A correlation between the mass of the most massive star and the
stellar density of companions is seen to exist around Herbig AeBe stars ({\sl Testi \etal},~1999)
although this is not necessarily incompatible with random sampling from
an IMF ({\sl Bonnell and Clarke}, 1999). Recently, {\sl Weidner and Kroupa}, (2006) have
suggested that observations indicate a strong correlation between 
the most massive star and the cluster mass, and that a random sampling
model can be excluded.  Estimates of  the number of
truly isolated massive stars are of order 4 \% or less  ({\sl de Wit \etal},~2005).
It is therefore a necessary condition for any model for the IMF to explain how
massive star formation occurs preferentially in the cores (see below) of stellar clusters where
stars are most crowded.

\subsection{Mass segregation}
Observations  show that young  stellar clusters generally have a significant 
degree of mass segregation with
the most massive stars located in the dense core of the cluster ({\sl Hillenbrand and Hartmann}, 1998;
{\sl Carpenter \etal},~2000; {\sl Garcia and Mermilliod}, 2001). For example, mass segregation is present in the 
Orion Nebula Cluster (ONC)
where stars more massive than five solar masses are significantly  more concentrated 
in the cluster core than are lower-mass stars ({\sl Hillenbrand and Hartmann}, 1988). 
This suggests that either the higher-mass stars
formed in the centre of the clusters, or that they moved there since their formation. 
Massive stars are expected to sink to the centre of the cluster due to two-body relaxation
but this dynamical relaxation occurs on the relaxation time, inversely proportional
to the stellar mass (e.g., {\sl Binney and Tremaine}, 1987). 
The young stellar clusters considered are generally less than a relaxation time old
such that dynamical mass segregation cannot have fully occurred.  N-body simulations
have shown that while some dynamical mass segregation does occur relatively quickly,
especially of the most massive star, the degree of mass segregation present 
cannot be fully attributed to dynamical relaxation. Instead, the mass segregation is 
at least partially primordial ({\sl Bonnell and Davies}, 1998; {\sl Littlefair \etal},~2003). 
 
 For example, in the ONC at 1 million years, the massive stars need to have formed
 within three core radii for two-body relaxation to be able to produce the central grouping 
of massive stars known as the Trapezium ({\sl Bonnell and Davies}, 1998).  Putting the massive stars 
at radii
 greater than the half-mass radius of the cluster 
implies that the ONC would have to be at least 10 dynamical times old 
(3 to 5 million years) in order to have a 20 \% chance of creating a Trapezium-like
 system in the centre due to dynamical mass segregation.  
It therefore appears to be an unavoidable consequence of star formation that higher-mass
stars typically form in the centre of stellar clusters. 
A caveat is that these conclusions
 depends on estimates of the stellar ages. If the systems are significantly older 
than is generally believed ({\sl Palla \etal},~2005), then dynamical relaxation is more likely to have contributed 
to the current mass segregation.

\subsection{Binary systems}

We know that many stars form in binary systems and that the binary
frequency increases with stellar mass. Thus, the formation of binary stars
is an essential test for models of the IMF. While the frequency of binaries
amongst the lower-mass stars and brown dwarfs is  $\approx$10-30 \% 
(see chapter by {\sl Burgasser \etal}),
this frequency increases to $\ge 50$ \%  for solar-type stars ({\sl Duquennoy and Mayor}, 1991)
and up to near 100
\% for massive stars ({\sl Mason \etal},~1998; {\sl Preibisch \etal},~2001; {\sl Garcia and Mermilliod}, 2001). 

Of added importance is that many of these systems are very close, with
separations less than the expected Jeans or fragmentation lengths within molecular clouds. 
This implies that they could not have formed at their present separations and masses but must
have either evolved to smaller separations, higher masses or both. An evolution in
binary separation, combined with a  continuum of massive
binary systems with decreasing separation down to a few stellar radii implies that
the likelihood for binary mergers should be significant. System mass ratios also
probably depend on primary mass as high-mass stars appear to have
an overabundance of similar mass companions relative to solar-type stars ({\sl Mason \etal},~1998; {\sl Zinnecker}, 2003).

The fact that binary properties (frequency, separations, mass ratios) depend
on the primary mass is important in terms of models for the IMF. 
Fragmentation
is unlikely to be able to account for the increased tendency of high-mass
binaries to have smaller separations and more similar masses relative
to lower-mass stars, whereas subsequent accretion potentially can ({\sl Bate and Bonnell}, 1997).

Understanding the binary properties, and how they depend on primary mass,  is also crucial in determining the IMF. 
For example, are the two components paired randomly or are they correlated in mass?
One needs to correct for unresolved binary systems and this requires detailed
knowledge of the distribution of mass ratios ({\sl Sagar and Richtler}, 1991; {\sl Malkov and Zinnecker}, 2001; {\sl Kroupa}, 2001). 

\section{NUMERICAL SIMULATIONS}

While numerical simulations provide a useful tool to test how the individual physical
processes affect the star formation and resulting initial mass function, it should be
recalled that each simulation  has its particular strengths
and weaknesses and that no simulation to date has included all the relevant physical
processes.  Therefore all conclusions based on numerical simulations should be 
qualified by the physics they include and their abilities to follow the processes involved.

The majority of the simulations used to study the origin of the IMF have used either grid-based
methods or the particle-based Smoothed Particle Hydrodynamics (SPH).  Grid based
codes use either a fixed Eulerian grid or an Adaptive Grid Refinement (AMR). 
Adaptive grids are a very important development as they provide much higher resolution
in regions of high density or otherwise of interest. This allows grid-based methods to follow
collapsing objects over many orders of magnitude increase in density. The  resolution
elements are individual cells although at least 8 cells are required in order to resolve
a 3-d self-gravitating object.
Grid-based
methods are well suited for including additional physics such as magnetic fields and radiation
transport. They are also generally better at capturing shocks as exact solutions across neighbouring
grid cells are straightforward to calculate. The greatest weaknesses of 
grid-based methods is the necessary advection of fluid through the grid cell, 
especially when considering a self-gravitating fluid.
Recently, {\sl Edgar \etal},~(2005) have shown how a resolved self-gravitating binary system can lose
angular momentum as it rotates through an AMR grid forcing the system to merge artificially. 

In contrast, SPH uses particles to sample the fluid and a smoothing kernel with which to
establish the local hydrodynamical quantities. The resolution element is the smoothing length
which generally contains $\approx$ 50 individual particles, but this is also sufficient
to resolve a self-gravitating object. Additionally, when following accretion flows, 
individual particles can be accreted.
SPH's primary asset is that it is Lagrangian and thus is ideally suited to follow
the flow of self-gravitating fluids. Gravity is calculated directly from the particles
such that it can easily follow a collapsing object. Following fragmentation requires
resolving the Jeans mass ({\sl Bate and Burkert },1997) at all points during the collapse. 
When the Jeans mass is not adequately resolved, fragments with masses below the resolution limit cannot be followed and are forced to disperse into the larger-scale environment.
This results in the simulation only determining a lower-limit
to the total number of physical fragments which should form. 
For the same reason, SPH cannot overestimate the number of fragments
that form.  Tests have repeatedly shown that artificial fragmentation does not  occur   
in SPH ({\sl Hubber \etal}, 2006; {\sl Bonnell and Bate}, in preparation), 
as any clumps that contain less than the minimum number of particles cannot collapse, 
be they gravitationally bound or not. Young stellar
objects can be represented by sink-particles that accrete all gas that flows into
their sink radius, and are bound to the star ({\sl Bate \etal},~1995). This permits simulations to follow the dynamics
much longer than otherwise possible and follow the accretion of mass onto individual stars.
It does exclude the possibility to resolve any discs interior to the sink-radius or their susbequent
fragmentation. 

Complicated fluid configurations such as occur in a turbulent medium are straightforward to follow due to the Lagrangian nature of the SPH method.
Including radiative transfer and magnetic fields are more complicated
due to the disorder inherent in a particle-based code. SPH also smoothes out shock fronts over
at least one kernel smoothing length,  but generally SPH does an adequate job of establishing the physical conditions across the shock. 

\begin{figure*}[t]
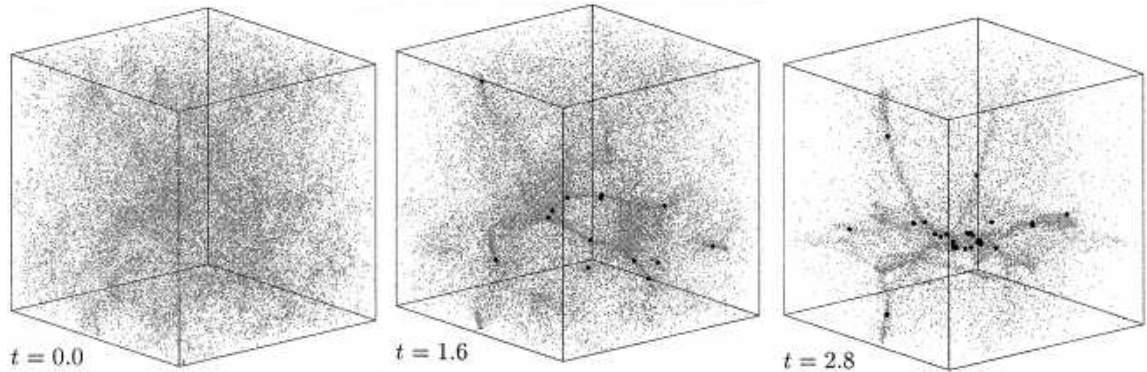

\epsscale{0.6}
\centerline{\plotone{bonnellOIMFfig2a.epsf}\plotone{bonnellOIMFfig2b.epsf}\plotone{bonnellOIMFfig2c.epsf}}
\caption{\label{klessenfrag} \small The gravitational fragmentation of molecular cloud is 
shown from a simulation containing initial structure ({\sl Klessen \etal},~1998). 
The gravitational collapse enhances this structure producing filaments which fragment 
to form individual stars. The time t is given in units of the free-fall time. }  
\end{figure*}

SPH's Lagrangian nature also makes it possible to trace
individual fluid elements throughout a simulation in order to establish what exactly is occurring,
something that is impossible with grid-based methods.
Furthermore, stringent tests
can be made on individual particles in order to avoid
unphysical results. For example, in the classical Bondi-Hoyle accretion flows, it is necessary
to resolve down below the Bondi-Hoyle radius in order to resolve the shock which allows the
gas to become bound to the star. This is a grave concern in grid-based codes 
as otherwise the accretion can be overestimated, but is less
of a worry in SPH as particles can be required to be bound before accretion occurs, even
when inside the sink-radius of the accretor. This ensures that the accretion is not
overestimated but under-resolved flows could result in an underestimation of the accretion rates. Particles
that would shock, become bound and accreted are instead free to escape the star.

\section{\textbf{PHYSICAL PROCESSES}}

\vspace{0.1cm} 

There are a number of physical processes which are likely to play an important
role in the star formation process and thus affect the resulting distribution of
stellar masses. These include gravity, accretion, turbulence, magnetic fields,
feedback from young stars and other semi-random processes such as dynamical ejections.

\subsection{Gravitational fragmentation}

It is clear that gravity has to play an important and potentially
dominant role in determining the stellar masses. Gravity is the one force
which we know plays the most important role in star formation, forcing molecular clouds
with densities of order $10^{-20}$ g  cm$^{-3}$ to collapse to form stars with densities
of order $1$ g cm$^{-3}$. It is therefore likely that gravity likewise plays
a dominant role in shaping the IMF.

Gravitational fragmentation is simply the tendency for gravity to generate clumpy structure
from an otherwise smooth medium. It occurs when a subpart of the medium is self-gravitating, 
that is when
gravitational attraction dominate over all support mechanisms. In astrophysics, the one support
that cannot be removed and is intrinsically isotropic (such that it supports an object 
in three dimensions) is the thermal pressure of the gas. Thus thermal
support sets a minimum scale on which gravitational fragmentation can occur.
The Jeans mass, 
based on the mass necessary for an object
to be bound gravitationally against its thermal support, can
be estimated  by comparing the respective energies and requiring 
that $|E_{\rm grav}| \ge \, E_{\rm therm}$. 
For the simplest case of a uniform density sphere this yields
\begin{equation}
\mj \approx 1.1  \left(T_{10}\right)^{3/2} \left(\rho_{19} \right)^{-1/2} \solm,  
\end{equation}
 where $\rho_{19}$ is the gas density in units of $10^{-19}$ g cm$^{-3}$ and $T_{10}$ 
is the temperature
 in units of 10 K. If external pressure is important, then one must use the Bonnor-Ebert
 mass ({\sl Ebert}, 1955; {\sl Bonnor}, 1956) which is somewhat smaller. 
 The corresponding Jeans length, or minimum length scale for gravitational fragmentation 
 is given by 
\begin{equation}
\rj \approx 0.057  \left(T_{10}\right)^{1/2} \left(\rho_{19} \right)^{-1/2} {\rm pc}.
\end{equation}
This gives an estimate of the minimum initial separation for self-gravitating fragments.

One can see that by varying the temperature and/or the density, it is straightforward
to obtain the full range of Jeans masses and thus potentially stellar masses 
and therefore a variation in either of these
variables can produce an IMF.  Generally, the temperature is  low before
star formation and assumed to be nearly isothermal
at $\approx 10K$ such that it is the density which primarily determines the Jeans mass.
Other forms of support, such as turbulence and magnetic fields have often been invoked
to set the Jeans mass (e.g., {\sl McKee and Tan}, 2003), but their relevance to gravitational
fragmentation is doubtful due to their non-isotropic nature.

\begin{figure}[h]
\epsscale{0.75}
\centerline{\plotone{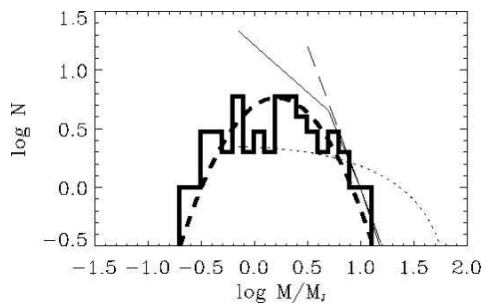}}
\caption{\label{klessenIMF} \small The IMF that results from isothermal
gravitational fragmentation 
(e.g., Fig.~\ref{klessenfrag}) is
typically broad and log-normal in shape ({\sl Klessen \etal},~1998). The stellar masses are
measured in terms of the average {\it initial} Jeans mass of the cloud.}  
\end{figure}

\begin{figure}[h]
\epsscale{0.5}
\centerline{\plotone{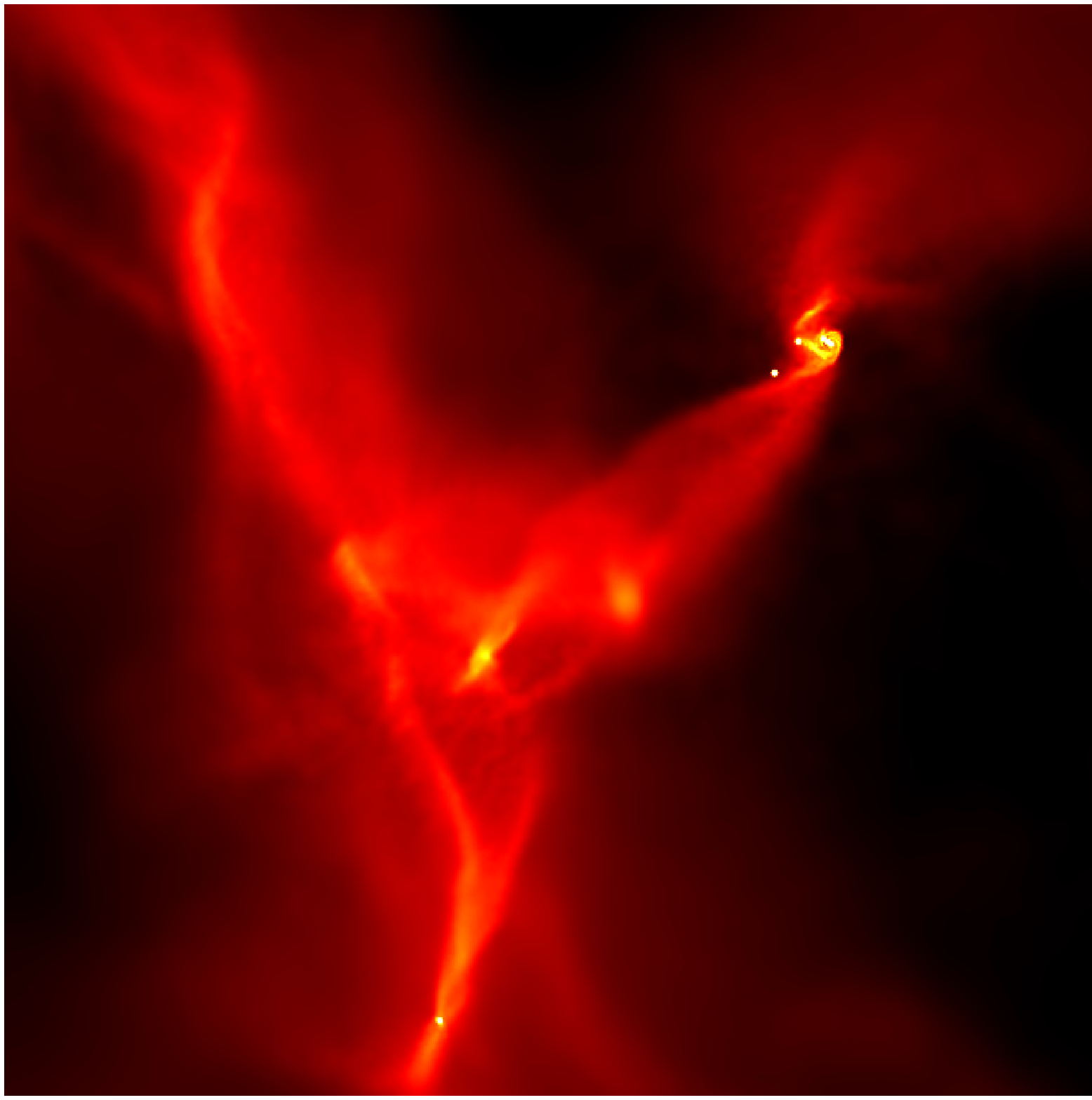}\plotone{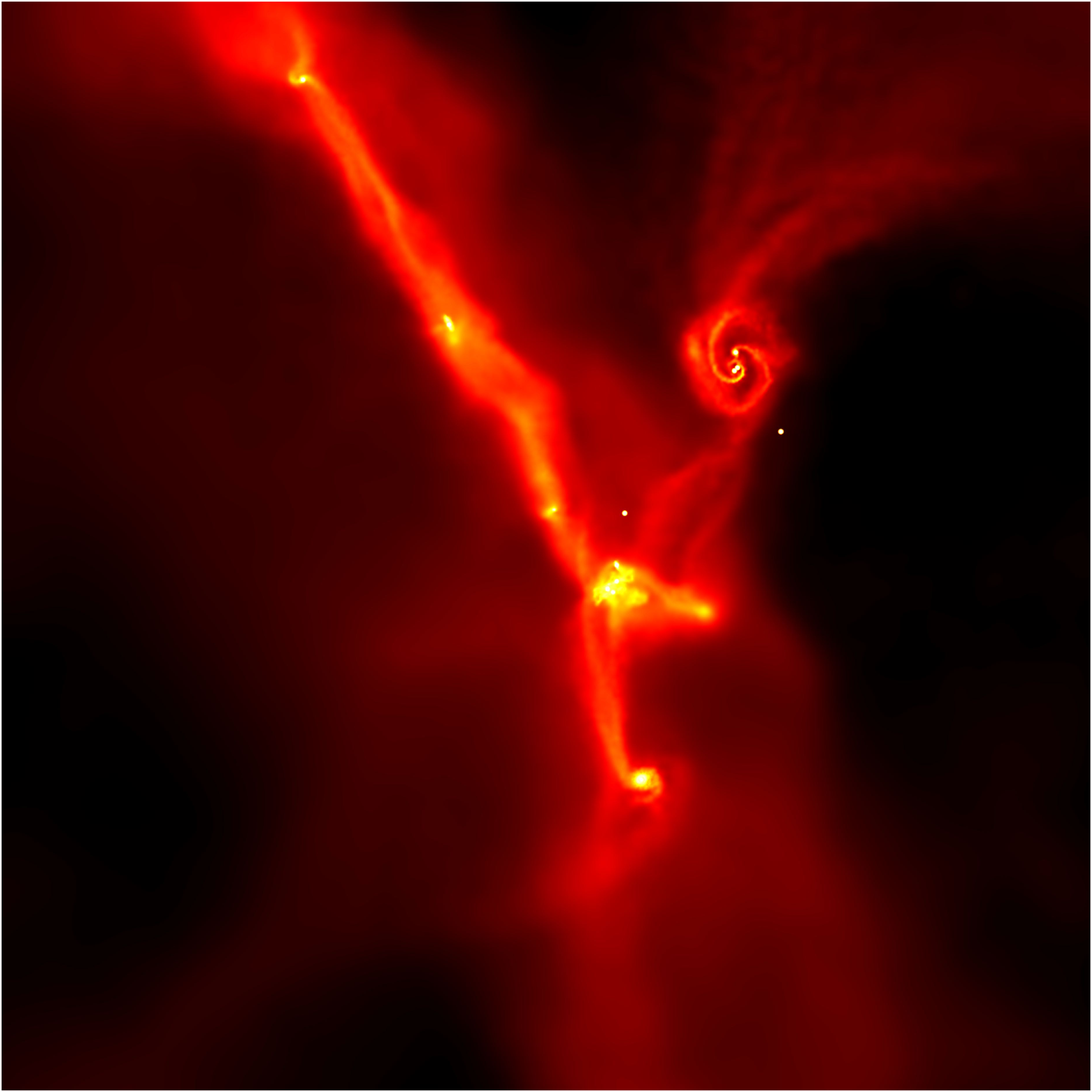}}
\caption{\label{batefrag} \small The fragmentation of filamentary structure, 
and the formation of low-mass stars and brown dwarfs, is shown in a simulation
of the formation of a small stellar cluster ({\sl Bate \etal},~2003).}  
\end{figure}

The primary requirement for gravitational fragmentation is that there exists
sufficient initial structure to provide a focus for the gravity. In a smooth uniform sphere, 
even if subregions are gravitationally unstable they will all collapse and merge together
at the centre of the cloud ({\sl Layzer}, 1963).
 Some form of seeding is required such that the local
free-fall time
\begin{equation}
\tff = \left(\frac{3 \pi}{32 G \rho}\right)^{1/2}
\end{equation}
is shorter than the global free-fall time of the cloud. In principle, any small perturbations
can be sufficient as long as the gas remains nearly isothermal ({\sl Silk}, 1982), but fragmentation
is effectively halted once the gas becomes optically thick and the collapse slows down 
({\sl Tohline}, 1982; {\sl Boss}, 1986).
Simple deformations in the form
of sheets ({\sl Larson}, 1985; {\sl Burkert and Hartmann}, 2004) or filaments ({\sl Larson}, 1985; {\sl Bastien \etal},~1991) 
are unstable to fragmentation 
as the local free-fall time is much shorter than that for the object as a whole. This
allows any density perturbations to grow non-linear during the collapse.  
More complex configurations (see Figure~\ref{klessenfrag})
 in the initial density field can equally result 
in gravitational fragmentation (e.g., {\sl Elmegreen and Falgarone}, 1996; {\sl Elmegreen}, 1997,~1999; 
{\sl Klessen \etal},~1998; {\sl Klessen and Burkert}, 2000,  2001). 
The origin of such density fluctuations
can be due to the 'turbulent' velocity field seen in molecular clouds (see below) providing the
seeds for the gravitational fragmentation ({\sl Klessen}, 2001; {\sl Bate \etal},~2003; {\sl Bonnell \etal},~2003,~2006a).

One of the general outcomes of gravitational fragmentation is that an upper limit 
to the number of fragments is approximately given by the number of Jeans masses present 
in the cloud ({\sl Larson},~1978,~1985; {\sl Bastien \etal},~1991;  {\sl Klessen \etal},~1998, {\sl Bate \etal},~2003). 
This is easily understood
as being the number of individual elements within the cloud that can be gravitationally bound. 
This  results in an average
fragment mass that is of order the Jeans mass {\sl at the time of fragmentation} (e.g., 
{\sl Klessen \etal},~1998; {\sl Klessen}, 2001; {\sl Bonnell \etal},~2004,~2006a; {\sl Clark and Bonnell} ,2005;  {\sl Jappsen \etal},~2005). The Jeans criterion can then be thought of as a
criterion to determine the characteristic stellar mass and thus provides the foundations for the
origin of the IMF.  Resulting IMFs are log-normal in shape (Fig.\ref{klessenIMF}, {\sl Klessen \etal},~1998; 
{\sl Klessen and Burkert}, 2001; {\sl Klessen}, 2001; {\sl Bate \etal},~2003). 
The problem is then what determines the Jeans mass at the point of fragmentation.
There are two possible solutions. 
First, that the initial conditions for star formation always have the same physical
conditions and thus the same Jeans mass of order a \solmas, which would seem unlikely. The second solution requires  some additional thermal physics which sets the Jeans mass at the point where
fragmentation occurs
({\sl Larson,} 2005, {\sl Spaans and Silk,}~2000). 

The coupling of gas to dust may provide the necessary physics to change from a cooling
equation of state ($T\propto \rho^{-0.25}$) to one including a slight heating 
($T\propto \rho^{0.1}$) with increasing gas densities ({\sl Larson}, 2005). 
This provides a method of setting the characteristic stellar mass which is then {\it independent
of the initial conditions for star formation}. Numerical simulations
using a simple cooling/heating prescription to mimic the effects
of this transition show that this sets the fragment mass and thus the peak
of the IMF ({\sl Jappsen \etal},~2005; {\sl Bonnell \etal},~2006a). Indeed, starting from
initial conditions with a Jeans mass of $5\solm$, which in an isothermal simulation
provide a nearly flat (in log mass) IMF up to $\approx 5 \solm$, the cooling/heating
equation of state reduces this characteristic mass to below $1\solm$ ({\sl Jappsen \etal},~2005;
{\sl Bonnell \etal},~2006a), allowing
for an upper-mass Salpeter-like slope due to subsequent accretion (see below).

Stellar masses significantly lower than the characteristic stellar mass are also 
explainable through gravitational fragmentation. In a collapsing region the gas density
can increase dramatically and this decreases the Jeans mass. The growth of filamentary
structure in the collapse (see Fig.~\ref{batefrag}), due to the funneling 
of gas into local potential minima, 
can then provide the seeds for fragmentation to form very low-mass objects such as brown dwarfs
({\sl Bate \etal},~2002a). Dense circumstellar discs also provide
the necessary low Jeans mass in order to form low-mass stars and brown dwarfs. 
Numerical simulations of gravitational fragmentation can thus
explain the characteristic stellar mass and the roughly flat 
(in log space) distribution of lower-mass 
stars and brown dwarfs ({\sl Bate \etal},~2003).

Gravitational fragmentation is unlikely to determine the full mass spectrum. 
It is difficult to see how gravitational fragmentation could account
for the higher-mass stars. These stars are born
in the dense cores of stellar clusters where stars are fairly closely packed. Their
separations can be used to limit the sizes of any pre-stellar fragments 
via the Jeans radius,
the minimum radius for an object to be gravitationally bound. This, combined with
probable gas temperatures, imply a high gas density and thus a low Jeans mass
({\sl Zinnecker \etal},~1993).
Thus, naively, it is low-mass and not high-mass stars that would be expected from a gravitational
fragmentation in the cores of clusters. In general, gravitational fragmentation would be expected to
instill a reverse mass segregation, the opposite of which is seen in young clusters.
Similarly, although fragmentation is likely to be responsible  for the formation of most binary
stars, it cannot explain the closest systems nor the tendency of higher-mass stars to be
in close systems with comparable mass companions.
 
\vspace{0.1cm} 
\subsection{Turbulence}
\vspace{0.1cm} 

It has long been known that supersonic motions are contained within molecular clouds (Larson 1981).
These motions are generally considered as being turbulent principally because of the linewidth-size
relation $ \sigma \propto R^{0.5}$ ({\sl Larson,}~1981; {\sl Heyer and Brunt},~2004) that mimics the expectation 
for turbulence ({\sl Elmegreen and Scalo},~2004) and implies an energy cascade from large to small scales.
Alternatively, the clouds could simply contain random bulk motions generated at all scales such as
occurs in a clumpy shock  ({\sl Bonnell \etal},~2006b).
Nevertheless, for the purposes of this review, we define turbulence as supersonic irregular
motions in the clouds that contribute to the support of these clouds 
(see chapter by Ballesteros-Paredes \etal). It is well known that
turbulence or its equivalent can generate density structures in molecular
clouds due to supersonic shocks that compress the gas 
({\sl Elmegreen},~1993; {\sl Vazquez-Semadeni},~1994; {\sl Padoan},~1995; {\sl Stone \etal},~1998; {\sl Mac Low \etal},~1998;
{\sl Ostriker \etal},~1999; {\sl Mac~Low and Klessen}, 2004; {\sl Elmegreen and Scalo}, 2004). 
The resultant
distribution of density structures, generally referred to as turbulent fragmentation, can either
provide the seeds for a gravitational fragmentation (e.g., references above, especially 
{\sl Mac Low and Klessen}, 2004),
or alternatively could determine the IMF directly at the pre-stellar core
phase of star formation ({\sl Padoan \etal},~1997,~2001; 
{\sl Padoan and Nordlund}, 2002).

\begin{figure}[h]
\epsscale{0.8}
\centerline{\plotone{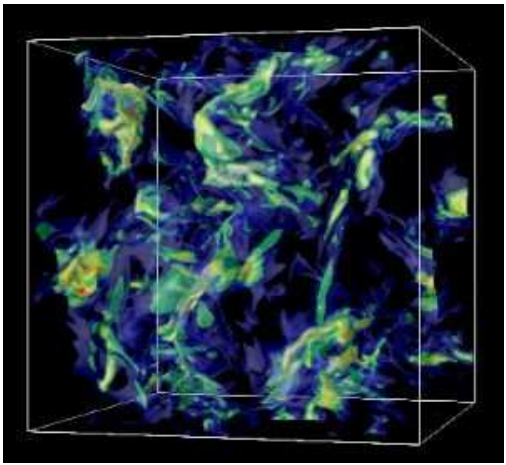}}
\caption{\label{bp2006turbfrag} \small The fragmentation of a turbulent medium
and the formation of prestellar clumps ({\sl Ballesteros-Paredes \etal},~2006a).}  
\end{figure}

Turbulent fragmentation provides an attractive mechanism to explain the IMF as it involves 
only one physical
process, which is observed to be ubiquitous in molecular clouds 
({\sl Elmegreen}~1993; {\sl Padoan \etal},~1997; 
{\sl Padoan and Nordlund},~2002). 
Multiple compressions result in the formation of sheets and then filaments
in the cloud. The density $\rho$ and widths $w$ of these filaments are due 
to the (MHD) shock conditions
such that higher Mach number shocks produce higher density but thinner filaments 
({\sl Padoan and Nordlund},~2002).  Clump masses can then be derived assuming that the 
shock width gives the
three-dimensional size of the clump ($M\propto \rho w^3$). 
High velocity shocks produce
high density but small clumps, and thus the lowest mass objects. 
In contrast,  lower velocity
shocks produce low-density but large shocks which account for the higher-mass clumps. 
Using the power-spectrum of velocities  from numerical simulations of turbulence,
and estimates of the density, $\rho$, and width, $w$, of MHD shocks as a function of the flow speed,
{\sl Padoan and Nordlund},~(2002) derive a clump mass distribution for turbulent fragmentation.
The turbulent spectrum results in a 'universal' IMF slope which closely resembles the Salpeter
slope. At lower masses, consideration of the likelihood that these clumps are sufficiently dense
to be Jeans unstable produces a turnover and a log normal shape into the brown dwarf regime.
This is calculated from the fraction of gas that is over the critical density 
for a particular Jeans mass, but it does not require that this gas is in a particular 
core of that mass.

\begin{figure}[h]
\epsscale{0.7}
\centerline{\plotone{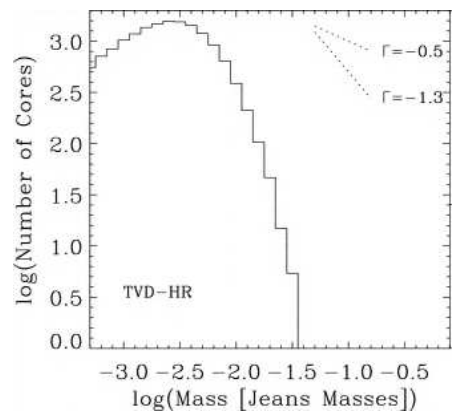}}
\caption{\label{bp2006turbcmd} \small The clump-mass distribution from a 
hydrodynamical simulation
of turbulent fragmentation ({\sl Ballesteros-Paredes \etal},~2006a). Note that the high-mass
end does not follow a Salpeter slope.}  
\end{figure}

Numerical simulations using grid-based codes have investigated the resulting clump-mass 
distribution from   turbulent fragmentation. 
While {\sl Padoan and Nordlund} (2004) have reported results consistent with their earlier 
analytical models, {\sl Ballesteros-Paredes \etal}~(2006a) conclude that the high-mass end of the
mass distribution is not truly Salpeter but becomes steeper at higher masses. Furthermore,
the shape depends on the Mach number of the turbulence implying that turbulent fragmentation
alone cannot reproduce the stellar IMF ({\sl Ballesteros-Paredes \etal},~2006a). The difference 
is attributed to having multiple shocks producing the density structure, which then
blurs the relation between the turbulent velocity spectrum and the resultant 
clump-mass distribution.  Thus, the higher mass clumps in the Padoan and Nordlund model
have internal motions which will sub-fragment them into smaller clumps 
(see chapter by {\sl Ballesteros-Parades \etal}).

The above grid-based simulations are generally not able
to follow any gravitational collapse and star formation so the question remains
open what stellar IMF would result.  SPH simulations that are capable of following the
gravitational collapse and star formation introduce a further complication. 
These simulations find that most of the clumps are generally unbound and therefore 
do not collapse to form stars ({\sl Klessen \etal},~2005; {\sl Clark and Bonnell}, 2005). 
It is only the most massive clumps that become gravitationally unstable and form stars. 
Gravitational collapse requires
masses of order the unperturbed Jeans  mass of the cloud suggesting that the
turbulence has played
only a minor role in triggering the star formation process ({\sl Clark and Bonnell}, 2005).
Even then, these
cores often contain multiple thermal Jeans masses and thus fragment to form several stars. 

In terms of observable predictions, the {\sl Padoan and Nordlund} turbulent compression model 
suggests, as does gravitational fragmentation,
that the minimum clump separations scales with the mass of the core. Thus, lower-mass
clumps can be closely packed whereas higher-mass cores need to be well separated.
If these clumps translate directly into stars as required for turbulent compression 
to generate the IMF,
then this appears to predict an initial configuration where the more massive stars are
in the least crowded locations. Unless they can dynamically migrate to the cores
of stellar clusters fairly quickly, then their formation is difficult to attribute 
to turbulent fragmentation.

Turbulence has also been invoked as a support for massive cores ({\sl McKee and Tan},~2003)
and thus as a potential source for massive stars in the centre of clusters.
The main idea is that the turbulence acts as a substitute for thermal support
and the massive clump evolves as if it was very warm and thus has a much higher Jeans mass.
The difficulty with this is that turbulence drives structures into objects
and therefore any turbulently supported clump is liable to fragment,
forming a small stellar cluster instead of one star. SPH simulations
have shown that, in the absence of magnetic fields, a centrally condensed
turbulent core fragments readily into multiple objects ({\sl Dobbs \etal},~2005). The fragmentation
is somewhat suppressed if the gas is already optically thick and thus non-isothermal. 
Heating from accretion onto a stellar surface can also potentially limit any fragmentation
({\sl Krumholz}, 2006)
but is likely to arise only after the fragmentation has occurred.
In fact, the difficulty really 
lies in how such a massive turbulent core could form in the first place. In a turbulent cloud, 
cores form  and dissipate on dynamical timescales suggesting that forming a long-lived core is
problematic ({\sl Ballesteros-Paredes \etal},~1999; {\sl Vazquez-Semadeni \etal},~2005).
As long as the region contains supersonic turbulence, it should fragment on its dynamical
timescale long before it can collapse as a single entity. Even MHD turbulence
does not suppress the generation of structures which will form the seeds for fragmentation
(see chapter by {\sl Ballesteros-Paredes \etal}).

The most probable role for turbulence is as a means for generating structure in
molecular clouds. This structure then provides the finite amplitude 
seeds for gravitational fragmentation
to occur, while the stellar masses are set by the local density and thermal properties
of the shocked gas. The formation of lower-mass
stars and brown dwarfs directly from turbulent compression is still an open question as it is 
unclear if turbulent compression can form gravitationally bound cores at such low masses. 
Turbulent compression is least likely to be responsible for  the high-mass slope of the IMF as 
numerical simulations suggest that the high-end core-mass distribution is not universal 
and does not follow a Salpeter-like slope (see Fig.~\ref{bp2006turbcmd}).
\vspace{0.1cm} 
\subsection{Accretion}
\vspace{0.1cm} 

\begin{figure}[h]
\epsscale{0.9}
\plotone{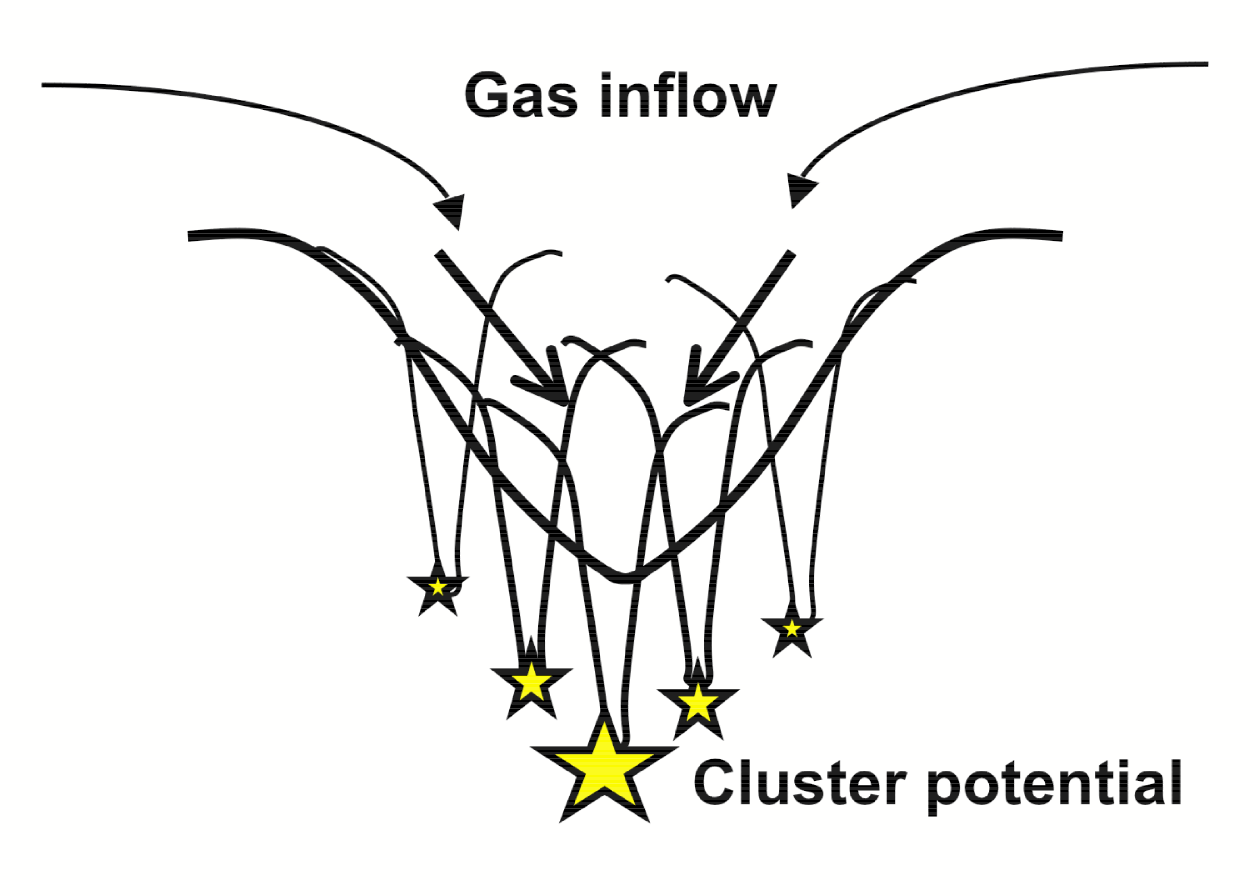}
\caption{\label{compacc} \small A schematic diagram of the physics of accretion 
in a stellar cluster: The 
gravitational potential of the individual stars form a larger scale potential
that funnels gas down to the cluster core. The stars located there are therefore able
to accrete more gas and become higher-mass stars. The gas reservoir can be replenished by
infall into the large-scale cluster potential.}  
\end{figure}

Gas accretion is a major process that is likely to play an important role in determining
the spectrum of stellar masses. To see this, one needs to consider three facts. 
First, gravitational
collapse is highly non-homologous ({\sl Larson}, 1969) with only a fraction of  a stellar mass
reaching stellar densities at the end of a free-fall time. The vast majority of the eventual star
needs to be accreted over longer timescales. Secondly, fragmentation is highly inefficient
with only a small fraction of the total mass being initially incorporated 
into the self-gravitating fragments
({\sl Larson}, 1978; {\sl Bate \etal},~2003).
Thirdly, and most importantly, mm observations of molecular clouds show that even when 
significant structure is present, this structure only comprises a few percent of the mass 
available ({\sl Motte \etal},~1998;
{\sl Johnstone \etal},~2000). The great majority of the cloud mass is in a more distributed form at
lower column densities, as detected by extinction mapping ({\sl Johnstone \etal},~2004). 
Young stellar clusters are also seen to have 70 to 90 \% of their
total mass in the form of gas ({\sl Lada and Lada},~2003).
Thus,  a large gas reservoir  exists such that if accretion of this gas does occur, it is likely to be the dominant contributer to the final 
stellar masses and the IMF.

Models using accretion as the basis for the IMF rely essentially  on the equation
\begin{equation}
\ms = \macc t_{\rm acc}, 
\end{equation}
and by having a physical model to vary either the accretion rate $\macc$ or the
accretion timescale $t_{\rm acc}$, can easily generate a full distribution of stellar masses.
In fact, accretion can be an extremely complex time-dependent phenomenon (e.g., {\sl Schmeja and Klessen},~2004) and it may
occur in bursts suggesting that we should consider the above equation 
in terms of a mean accretion rate and timescale.
The accretion rates can be varied by being mass dependent 
({\sl Larson}, 1978; {\sl Zinnecker}, 1982; {\sl Bonnell \etal},~2001b), dependent
on variations of the gas density ({\sl Bonnell \etal,}~1997, 2001a) or 
on the relative velocity between gas and stars ({\sl Bondi and Hoyle}, 1944; {\sl Bate \etal},~2003). 
Variations in $t_{\rm acc}$  ({\sl Basu and Jones}, 2004; {\sl Bate and Bonnell}, 2005) can be due
to ejections in clusters ({\sl Bate \etal},~2002a, see \S 5.6 below) or feedback from forming stars 
({\sl Shu \etal},~2004; {\sl Dale \etal},~2005, see \S 5.5 below). 

The first models based on accretion ({\sl Larson},~1978,1982; {\sl Zinnecker},~1982) discussed how
stars compete from the available mass in a reservoir. Stars that accrete slightly more
due to their initial mass or proximity to more gas ({\sl Larson},~1992) increase their
gravitational attraction and therefore their ability to accrete. The depletion
of the gas reservoir means that there is less for the remaining stars to accrete.
This competitive accretion then
provides a reason why there are a few high-mass stars compared to
a much larger number of low-mass stars.  

\begin{figure*}[t]
\epsscale{2.0}
\plotone{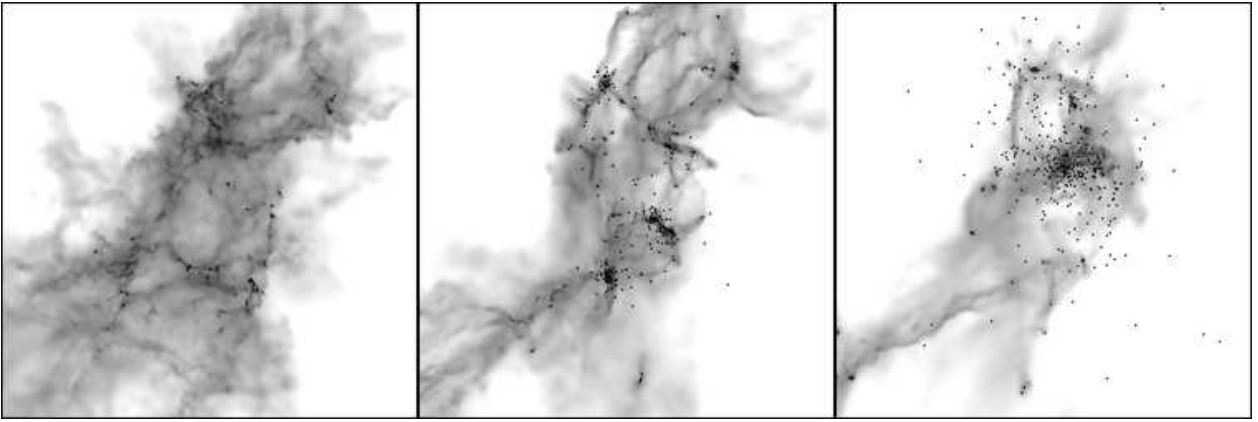}
\caption{\label{cluform} \small The fragmentation of a 1000\,$M_{\odot}$ 
turbulent molecular cloud
and the formation of a stellar cluster  ({\sl Bonnell \etal},~2003).
Note the merging of the smaller subclusters to a single big cluster.}  
\end{figure*}

In a stellar cluster, the accretion is complicated by the overall potential of the system. 
Figure~\ref{compacc} shows schematically the effect of the cluster potential
on the competitive accretion process.  The gravitational potential is the combined potential 
of all the stars
and gas contained in the cluster. This potential then acts to funnel gas down to the centre 
of the cluster
such that any stars located there have significantly higher accretion rates 
({\sl Bonnell \etal},~1997,~2001a).
These stars therefore have a greater ability to become higher-mass stars due to the
higher gas density and due to the fact that
this gas is constantly being replenished by infall from
the outer part of the cluster.  
Stars that accrete more are also more liable to sink to the centre of the potential
and thereby increase their accretion rates further.
It is worthwhile noting here that this process would occur 
even for a static potential where the stars do not move. 
The gas is being drawn down to the centre of the
potential. It has to settle somewhere, and unless it is already a self-gravitating fragment 
(ie a protostar),
it will fall into the local potential of one of the stars. 

Stars not in the centre of the cluster accrete less as gas is spirited away 
towards the cluster centre.
This ensures that the mean stellar mass remains close to the characteristic mass given
by the fragmentation process.  Accretion rates onto individual stars depend on 
 the local
gas density, the mass of the star and the relative velocity between the gas and the star:
\begin{equation}
\macc \approx \pi \rho \vrel \racc^2,
\end{equation}
where $\racc$ is the accretion radius which depends on the mass of the star (see below). 
The accretion radius is the radius at which gas is irrevocably
bound to the star. As a cautionary note,  in a stellar cluster  the local gas density depends
on the cluster potential and the relative  gas velocity can be very different from 
the star's velocity in the rest frame of the cluster,  as both gas and stars are experiencing the same accelerations.

Numerical simulations ({\sl Bonnell \etal},~2001a) show that in a stellar cluster the
accretion radius depends on whether the gas or the stars dominate the potential. In the former
case, the relative velocity is low and accretion is limited by the star's tidal radius. 
This is given by
\begin{equation}
\Rtidal \approx 0.5  \left({\ms / \menc}\right)^{1\over 3} \Rclus,
\end{equation}
which measures at what distance gas
is more bound to an individual star rather than being tidally sheared away by the overall 
cluster potential. The tidal radius depends on the star's position in the cluster, via
the enclosed cluster mass $\menc$ at the radial location of the star $\Rclus$.
The alternative is if the stars dominate the potential,
then the relative velocity between the gas and the stars can be high.  
The accretion radius is then the
more traditional Bondi-Hoyle radius of the form
\begin{equation}
\Rbh \approx 2G\ms/(\vrel^2 + c_s^2).
\end{equation}
It is always the smaller of these two accretion radii which 
determines when gas is bound to the star and
thus should be used to determine the accretion rates. We note again that the
relative gas velocity can differ significantly from the star's velocity in the rest
frame of the cluster.  
Using a simple model for a stellar cluster, it is straightforward to show that these two
physical regimes result in two different IMF slopes because of the differing mass dependencies
in the accretion rates ({\sl Bonnell \etal},~2001b). 
The tidal radius accretion has $\macc \propto \ms^{2/3}$ and, in a $n \propto r^{-2}$
stellar density distribution  results
in a relatively shallow $dN \propto \ms^{-1.5} d\ms$ (cf. {\sl Klessen and Burkert}, 2000). 
Shallower stellar density distributions produce steeper IMFs. For accretion
in a stellar dominated potential, 
Bondi-Hoyle accretion 
in a uniform gas distribution results in an IMF of the form $dN \propto \ms^{-2} d\ms$
({\sl Zinnecker},~1982). To see this, consider an accretion rate based on equations (8) and (10) 
\begin{equation}
\macc \propto \ms^2,
\end{equation}
with a solution 
\begin{equation}
\label{maccbhsol}
\ms = {\msi \over 1 - \beta \msi t}
\end{equation}
where $\msi$ is the initial stellar mass and $\beta$ includes the 
dependence on gas density and velocity (assumed constant in time).
From equation(\ref{maccbhsol}) we can derive a mass function $dN =F(\ms)d\ms$ by noting that
there is a one to one mapping of the initial and final stellar masses (ie, that the
total number of stars is conserved and  that there is a monotonic relation between initial and final masses) such that  
\begin{equation}
\label{imfmap}
F(\ms) d\ms = F(\msi) d\msi.
\end{equation}
Using equation(\ref{maccbhsol})   we  can easily  
derive 
\begin{equation}
F(\ms) = F(\msi) \left({\ms / \msi}\right)^{-2}.
\end{equation}
which, in the case where there is only a small range of initial stellar masses and for  $\ms >> \msi$,
gives
\begin{equation}
dN \propto \ms^{-2} d\ms,
\end{equation}
whereas if the 'initial'  mass distribution is initially significant and decreasing with
increasing masses, then the resulting IMF is steeper. 
In a stellar cluster with a degree of mass segregation
from an earlier gas-dominated phase, this results in a steeper IMF  closer to  
$dN \propto \ms^{-2.5} d\ms$ ({\sl Bonnell \etal},~2001b).
This steeper IMF is therefore appropriate for the more massive stars that 
form in the core of a cluster because
it is there that the stars first dominate the cluster potential. 
Although the above is a semi-analytical model and suffers from the pitfalls 
described in introduction, it is  comforting to note
that numerical simulations do reproduce the above IMFs and additionally show that the
higher-mass stars accrete the majority of their mass in the stellar dominated regime
which should, and in this case does,  produce the steeper Salpeter-like IMF ({\sl Bonnell \etal},~2001b).

\begin{figure}[h]
\epsscale{0.6}
\plotone{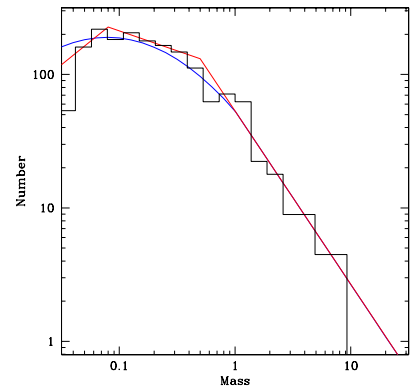}
\caption{\label{simIMF} \small The resulting IMF from a simulation of the fragmentation and
competitive accretion in a forming stellar cluster (e.g., {\sl Bonnell \etal},~2003) is shown 
as a function of log mass ($dN({\rm log\ mass})$).
Overplotted is the three segment power-law IMF
from {\sl Kroupa} (2001) and Chabrier's (2003) log-normal plus power-law IMF. The masses from the simulation have been rescaled to
reflect an initial Jeans mass of $\approx 0.5$ \solmas.   }
\end{figure}

A recent numerical simulation showing the fragmentation of a turbulent molecular
cloud and the formation of a stellar cluster is shown in Figure\ref{cluform}. 
The newly formed stars fall into local potential minima, forming small-N systems
which subsequently merge to form one larger stellar cluster. 
The initial fragmentation produces objects with masses 
comparable to the mean Jeans mass
of the cloud ($\approx 0.5 \solm$) which implies that they are formed due to gravitational, not turbulent, fragmentation. It is  the  subsequent competitive
accretion which forms the higher-mass stars ({\sl Bonnell \etal},~2004) and 
thus the Salpeter-like power law part of the IMF. Overall, the simulation forms
a  complete stellar population that 
follows a realistic IMF from $0.1 \solm$ to $30 \solm$ (Fig.\ref{simIMF} and 
{\sl Bonnell \etal},~2003). 
Accretion forms six stars in excess of 10 \solmas\ with the most
massive star nearly 30 \solmas. Each forming sub-cluster contains a
 more massive star in its centre and has a population consistent
with a Salpeter IMF ({\sl Bonnell \etal},~2004).

\begin{figure}[h]
\epsscale{0.7}
\plotone{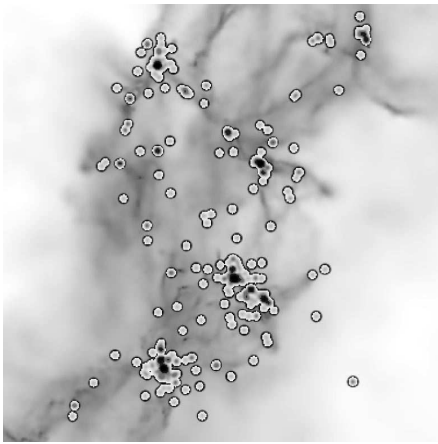}
\caption{\label{massseg} \small The location of the massive stars (dark circles) is shown to be
in the centre of individual  subclusters  of low-mass stars (light circles) 
due to competitive accretion (cf. {\sl Bonnell \etal},~2004).}  
\end{figure}

One of the advantages of such a model for the IMF is that it automatically results in a mass
segregated cluster. This can be seen from the schematic Figure~\ref{compacc} showing how
the stars that are located in the core of the cluster benefit from the extended cluster
potential to increase their accretion rates over what they would be in isolation. 
Thus  stars more massive than the mean stellar mass should be relatively mass segregated
from birth in the cluster. 
This is shown in Figure~\ref{massseg} which displays the distribution of low-mass
stars with the higher-mass stars located in the centre of individual clusters. There is always
a higher-mass star in every (sub)-cluster. Even when individual sub-clusters merge,
the massive stars quickly settle into the centre of the combined potential thereby benefitting
most from any continuing accretion. One of the strong predictions of competitive accretion is that
there is a direct correlation between the formation of a stellar cluster and the most massive star
it contains. Accretion and the growth of the cluster are linked such that the system always
has a realistic IMF. 

There have recently been some concerns raised that accretion cannot produce
the high-mass IMF either due to numerical reasons ({\sl Krumholz \etal},~2006) or
due to the turbulent velocity field ({\sl Krumholz \etal},~2005a). The numerical
concern is that SPH calculations may overestimate the accretion rates if they
do not resolve the Bondi-Hoyle radius. However, SPH simulations, being particle based, 
ensure that unphysical accretion does not occur by demanding that any gas
that is accreted is bound to the star. The second concern is that accretion rates
should be too low in a turbulent medium to affect the stellar masses. Unfortunately,
this study assumes that gravity
is negligible on large scales except as a boundary condition for the star forming
clump. This cannot be correct in a forming stellar cluster where both gas and stars undergo significant gravitational accelerations from the cluster potential. Furthermore, {\sl Krumholz \etal} take a virial velocity for the clump to use as the
turbulent velocity neglecting that turbulence follows a velocity-sizescale $v\propto R^{1/2}$
law ({\sl Larson}, 1981; {\sl Heyer and Brunt} ,2004). SPH simulations show that mass accretion
occurs from lower velocity gas initially, proceeding to higher velocities when the
stellar mass is larger, consistent with both the requirements of the turbulent scaling
laws and Bondi-Hoyle accretion ({\sl Bonnell and Bate},~2006, in preparation).

\vspace{0.1cm} 
\subsection{Magnetic Fields}
\vspace{0.1cm} 

Magnetic fields are commonly invoked as an important mechanism for star formation
and thus need to be considered as a potential mechanism for affecting the IMF. Magnetic
 fields were initially believed to dominate molecular clouds with ambipolar diffusion of these
fields driving the star formation process ({\sl Mestel and Spitzer}, 1956; {\sl Shu \etal},~1987). 
Since the realisation that
ambipolar diffusion takes too long, and that it would inhibit fragmentation and 
thus the formation of multiple
stars and clusters, and crucially that supersonic motions are common in molecular clouds,
the perceived role of magnetic fields has been revised to one of increasing 
the lifetime of turbulence
({\sl Arons and Max},~1975; {\sl Lizano and Shu},~1989).
More recently, it has been shown that magnetic fields have little effect on the decay rate of
turbulence as they do not fully cushion shocks ({\sl Mac Low \etal},~1997; {\sl Stone \etal},~1997). 
Still, magnetic fields
are likely to be generally present in molecular clouds and can play an important, if still
relatively unknown role. 

There have been many studies into the evolution of MHD turbulence and structure
formation in molecular clouds (e.g., {\sl Ostriker \etal},~1999, {\sl Vazquez-Semadeni \etal},~2000 ; 
{\sl Heitsch \etal},~2001; {\sl Tilley and Pudritz}, 2005;  {\sl Li and Nakamura},~2004; see chapter by 
{\sl Balesteros-Paredes \etal}).  
These simulations have found that both MHD and pure HD simulations result
in similar clump-mass distributions. One difference is that the slightly
weaker shocks in MHD turbulence shift the clump-masses to slightly higher masses.

One potential role for magnetic fields which has not been adequately explored is that they
could play an important role in setting the characteristic stellar mass in terms of an
effective  magnetic Jeans mass. 
Although in principle this is easy to derive, it is unclear how it would work in practice as magnetic
fields are intrinsically non-isotropic and therefore the analogy to an isotropic pressure support
is difficult to make. Recent work on this by {\sl Shu \etal}~(2004) has investigated whether
magnetic levitation, the support of the outer envelopes of collapsing cores, can set
the characteristic mass. Inclusion of such models into numerical simulations is needed
to verify if such processes do occur.

\vspace{0.1cm} 
\subsection{Feedback}
\vspace{0.1cm} 

Observations of star forming regions readily display the fact that young stars have
a significant effect on their environment. This feedback, including jets and outflows from
low-mass stars, and winds, ionisation, and radiation pressure from high-mass stars,
is therefore a good candidate to halt the accretion process and thereby set the stellar masses
({\sl Silk}, 1995, {\sl Adams and Fatuzzo}, 1996).
To date, it has been difficult to construct a detailed model for the IMF from feedback as it is
a rather complex process. Work is ongoing to include the effects
of feedback in numerical models of star formation but have not yet been able to generate 
stellar mass functions  ({\sl Li and Nakamura},~2005).  In these models, feedback injects
significant kinetic energy into the system which appears to quickly decay away again
({\sl Li and Nakamura},~2005). Overall, the system continues to evolve (collapse) in a similar way to 
simulations that neglect both feedback and magnetic fields (e.g., {\sl Bonnell \etal},~2003).

Nevertheless, we can perhaps garner some insight
from recent numerical simulations including the effects of ionisation from massive stars
({\sl Dale \etal},~2005). The inclusion of ionisation from an O star into a simulation of
the formation of a stellar cluster shows that the intrinsically isotropic radiation
escapes in preferential directions due to the non-uniform gas distributions
(see also {\sl Krumholz \etal},~2005b). Generally,
the radiation decreases the accretion rates but does not halt accretion. In more
extreme cases where the gas density is lower, the feedback halts the accretion
almost completely for the full cluster. This implies that feedback can stop accretion
but probably not differentially and therefore does not result in a non-uniform $t_{\rm acc}$
which can be combined with a uniform $\macc$ to form a stellar IMF.

Feedback from low-mass stars is less likely to play an important role in setting the IMF.
This is simply due to the well collimated outflows being able to deposit their energy at
large distances from the star forming environment ({\sl Stanke \etal},~2000).  As accretion can continue
in the much more hostile environment of a massive star where the feedback is intrinsically
isotropic, it is difficult to see a role for well collimated outflows in setting the IMF.

\vspace{0.1cm} 
\subsection{Stellar interactions}
\vspace{0.1cm} 

The fact that most stars form in groups and clusters, and on smaller scales in binary and multiple
systems, means that they are likely to interact with each other on timescales comparable
to that for gravitational collapse and accretion. 
By interactions, we generally mean gravitational interactions ({\sl Reipurth and Clarke}, 2001) 
although in the dense cores of stellar clusters this could 
involve collisions and mergers ({\sl Bonnell \etal},~1998, {\sl Bally and Zinnecker},~2005)  
These processes are essentially random with
a probability given by the stellar density, velocity dispersion and stellar mass. 
Close encounters with
binary or higher-order systems generally result in an exchange of energy which can eject
the lower-mass objects of the encounter ({\sl Reipurth and Clarke}, 2001). Such an event can quickly remove an accreting star
from its gas reservoir, thereby truncating its accretion and setting the stellar mass. This
process is what is seen to occur in numerical simulations of clustered star formation 
({\sl Bate \etal},~2002a,~2003) where low-mass objects are preferentially ejected. These objects are often
then limited to being brown dwarfs whereas they could have accreted up to stellar masses had
they remained in the star forming core (see also {\sl Price and Podsialowski},~1995).

 
Numerical simulations including the dynamics of the newly formed stars have repeatedly shown
that such interactions are relatively common ({\sl McDonald and Clarke},~1995; {\sl Bonnell \etal},~1997;
{\sl Sterzik and Durisen},~1998,~2003; {\sl Klessen and Burkert},~2001; {\sl Bate \etal},~2003, {\sl Bate and Bonnell},~2005), especially in small-N or subclusters
where the velocity dispersion is relatively low. Thus, such a mechanism should populate the
entire regime from the smallest Jeans mass formed from thermal (or turbulent) fragmentation
up to the characteristic mass. This results in  a relatively flat IMF (in log mass) for low-mass objects ({\sl Klessen and Burkert}, 2001; {\sl Bate \etal},~2003, {\sl Bate and Bonnell}, 2005; {\sl Bate}, 2005; {\sl Delgado-Donate \etal}, 2004).

Stellar mergers are another quasi-random event that could occur in very dense cores
of stellar clusters involving mergers of intermediate or high-mass single ({\sl Bonnell \etal},~1998; {\sl Bonnell and Bate}, 2002; {\sl Bally and Zinnecker}, 2005) or binary ({\sl Bonnell and Bate}, 2005) stars. In either case, mergers require relatively high stellar
densities of order $10^8$  and $10^6$ stars pc$^{-3}$ respectively. These densities, although higher than generally observed are conceivably due to a likely high density phase in the early
evolution of stellar clusters ({\sl Bonnell and Bate}, 2002; {\sl Bonnell \etal},~2003). In fact, estimates of
the resolved central stellar density in 30 Doradus and  the Arches cluster are of order a few  $\times 10^5$ \solmas\ pc$^{-3}$ ({\sl Hofmann \etal},~1995; {\sl Stolte \etal},~2002). Such events could play
an important role in setting the IMF for the most massive stars.

\vspace{0.1cm} 

\subsection{Summary of processes}

\begin{figure}[h]
\epsscale{0.8}
\plotone{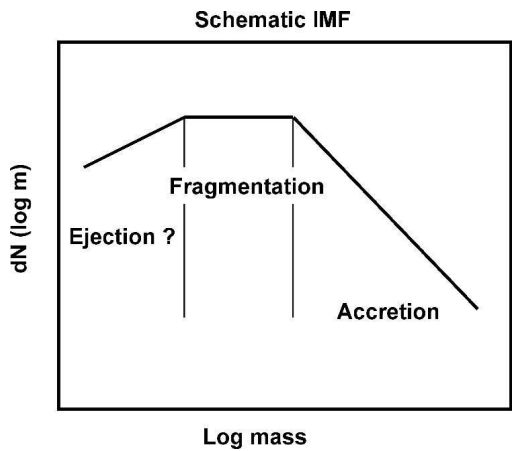}
\caption{\label{expIMF} \small A schematic IMF showing the regions that are expected to be
due to the individual processes. 
The peak of the IMF and the characteristic stellar mass are believed
to be due to gravitational fragmentation, while lower mass stars are best understood as being due
to fragmentation plus ejection or truncated accretion while higher-mass stars are understood
as being due to accretion.}  
\end{figure}

From the above  arguments  and the expectation of the different physical processes,
we can start to assess what determines the stellar masses in the various regimes
(Fig.~\ref{expIMF}). 
A general caveat should be  noted that we still do not have a thorough understanding
of what magnetic fields and feedback can do but it is worth noting that in their
absence we can construct a working model for the origin of the IMF. 
First of all, we conclude that the characteristic stellar mass and the broad peak
of the IMF is best attributed to gravitational fragmentation and the accompanying thermal
physics which sets the mean Jeans mass for fragmentation. The broad peak
can be understood as being due to the dispersion in gas densities and temperature
at the point where fragmentation occurs. Turbulence is a necessary
condition in that it generates the filamentary structure in the molecular clouds which
facilitates the fragmentation, but does not itself set the median or characteristic stellar mass. 

Lower-mass stars are
most likely formed through the gravitational
fragmentation of a collapsing region such that the increased gas density allows
for lower-mass fragments. These fragments  arise in collapsing
filaments and circumstellar discs ({\sl Bate \etal},~2002a). 
A crucial aspect of this mechanism for the formation of low-mass stars and brown dwarfs
is that they not be allowed to 
increase their mass significantly through accretion. If lower-mass stars
are indeed formed in gas dense environments to achieve the low Jeans masses, then
subsequent accretion can be expected to be significant. Their continued
low-mass status  requires that
they are ejected from their natal environment, or at least that they are accelerated
by stellar interactions such that their accretion rates drops to close to zero.
The turbulent compressional formation of low-mass
objects ({\sl Padoan and Nordlund}, 2004) is potentially a viable mechanism although 
conflicting simulations
have raised doubts as to whether low-mass {\sl gravitationally bound}
cores are produced which  then collapse to form stars.

Lastly, we conclude that the higher-mass IMF is  probably 
due to continued accretion in a clustered environment. 
A  turbulent compression origin for higher mass stars is problematic 
as the core-mass distribution from turbulence does not appear to be universal 
({\sl Ballesteros-Paredes \etal},~2006). Furthermore, the large sizes of higher mass prestellar
cores generated from turbulence suggest that they should be found in low stellar
density environments, not in the dense cores of stellar clusters. Nor should
they then be in close, or even relatively wide, binary systems.
In contrast, the ability of the cluster potential to increase accretion rates
onto the stars in the cluster centre is a simple explanation for more massive stars in the context
of low-mass star formation. Continuing accretion is most important for the more massive stars in a forming cluster because it is these that settle, and remain, in the denser
central regions. This also produces the observed mass segregation in
young stellar clusters.  The strong mass dependency of the accretion rates  ($\propto M^2$) 
results in  a  Salpeter-like high-mass IMF. 

Continued accretion and dynamical interactions can also potentially explain
the existence of closer binary stars, and the dependency of binary properties on stellar
masses  ({\sl Bate \etal},~2002b; {\sl Bonnell and Bate},~2005; 
{\sl Sterzik \etal},~2003; {\sl Durisen \etal},~2001).  
Dynamical
interactions harden any existing binary  ({\sl Sterzik and Durisen}, 1998,~2003; {\sl Kroupa}, 1995) 
and continued accretion increases
both the stellar masses and the binding energy of the system 
({\sl Bate and Bonnell}, 1997; {\sl Bonnell and Bate}, 2005). 
This can explain the higher frequency of binary systems amongst
massive stars, and the increased likelihood that these systems are close and of near-equal
masses.

\section{OUTSTANDING PROBLEMS}

There are outstanding issues which need to be resolved in order to fully
understand the origin of the IMF. Some of these involve detailed
understanding of the process (e.g., massive star formation, mass limits) whereas others 
include new observations which may be particularly useful in determining
the origin of the IMF. 

\subsection{Clump-mass spectrum}

In order for the stellar IMF  to come directly from the clump-mass spectra 
observed in molecular clouds (e.g., {\sl Motte \etal},~1998; {\sl Johnstone \etal},~2000), 
a one-to-one mapping of core clump to stellar mass is required. The
high frequency of multiple systems even amongst the youngest stars ({\sl Duch\^ene \etal},~2004)
makes a one-to-one mapping unlikely for masses near solar and above. 
At lower masses, the reduced frequency
of binary systems (e.g., {\sl Lada, 2006}) means that a one-to-one mapping is potentially viable.
Another potential difficulty is that some, especially lower mass, clumps are
likely to be transient ({\sl Johnstone \etal},~2000). Simulations commonly report
that much of the lower-mass structure formed is gravitationally unbound ({\sl Klessen}, 2001;
{\sl Clark and Bonnell},~2005; {\sl Tilley and Pudritz},~2005). 
Furthermore, as such mass spectra can be understood to arise due
to purely hydrodynamical effects without any self-gravity (e.g., {\sl Clark and Bonnell},~2006),
the relevance for star formation is unclear. 
If the clump-mass spectrum does play an integral role in the origin of the IMF, then
there should be additional evidence for this in terms of observational properties
that can be directly compared. For example, the clustering and spatial mass distribution
of clumps should compare directly and favorably to that of the youngest class 0 sources 
(e.g., {\sl Elmegreen and Krakowski},~2001).

\subsection{Massive stars}

The formation of massive stars, with masses in excess of $10 \solm$, is problematic
due to the high radiation pressure on dust grains and because of the dense
stellar environment in which they form. The former can actually halt the infall
of gas and thus appears to limit stellar masses. Simulations to date suggest
that this sets an upper-mass limit to accretion somewhere in the 10 to 40 \solmas\ range ({\sl Wolfire
and Casinelli},~1986;  {\sl Yorke and Sonnhalter},~2002; {\sl Edgar and Clarke},~2004). 
Clearly, there needs to be a mechanism for circumventing this problem as stars
as massive as 80-150\  \solmas\ exist ({\sl Massey and Hunter}, 1998; {\sl Weidner and Kroupa}, 2004; {\sl Figer}, 2005). Suggested solutions include
disc accretion and radiation beaming, ultra high accretion rates that overwhelm the radiation pressure
({\sl McKee and Tan} ,2003),
Rayleigh Taylor instabilities in the infalling gas ({\sl Krumholz \etal},~2005c) and stellar collisions ({\sl Bonnell \etal}, 1998; {\sl Bonnell and Bate},~2002,~2005). 
The most complete simulations of disk accretion ({\sl Yorke and Sonnhalter},~2002) suggest that radiation beaming
due to the star's rapid rotation, combined with disc accretion can reach stellar masses
of order 30 to 40 \solmas, although with low efficiencies.  What is most important for
any mechanism for massive star formation is that it be put into  the context of forming a full IMF
(e.g., {\sl Bonnell \etal},~2004). The most likely scenario for massive star formation involves
a combination of many of the above processes, competitive accretion in order to
set the distribution of stellar masses, disk accretion, radiation beaming and potentially
Rayleigh-Taylor instabilities or even stellar mergers to overcome the radiation pressure.
Any of these could result in a change in the slope
of high-mass stars reflecting the change in physics.

\subsection{Mass limits}

Observationally, it is unclear what limits there are on stellar masses. At low
masses, the IMF appears to continue as far down as is observable. Upper-mass limits are on firmer ground observationally
with strong evidence of a lack of stars higher than $\approx 150$ \solmas\ 
even in regions where statistically they are expected ({\sl Figer}, 2005; {\sl Oey and Clarke}, 2005;
{\sl Weidner and Kroupa},~2004). Physically, the only limitation on the formation
of low-mass objects is likely to be the opacity limit whereby an object
cannot cool faster than it contracts, setting a lower limit for a gravitationally
bound object ({\sl Low and Lynden-Bell}, 1976; {\sl Rees}, 1976; {\sl Boyd and Whitworth}, 2005).
This sets a minimum Jeans mass of order $3-10$ Jupiter masses.
At the higher-end physical limits could be set by radiation pressure on 
dust or electrons
(the Eddington limit), or by physical collisions.


\subsection{Clustering and the IMF}

Does the existence of a bound stellar cluster affect the high-mass end of the IMF? 
If accretion in a clustered
environment is responsible for the high-mass IMF, then there should be a direct
link between cluster properties and the presence of high-mass stars. 
Competitive accretion models require the presence of a stellar cluster in order 
for the distributed gas
to be sufficient to form high-mass stars. Thus, a large-N cluster produces
a more massive star than does the same number of stars divided into many
small-N systems (e.g., {\sl Weidner and Kroupa}, 2006). The combined number of stars in the small-N
systems should show a significant lack of higher-mass stars. 
Evidence for such an environmental dependence on
the IMF has recently been argued based on observations of the Vela D cloud ({\sl Massi \etal}, 2006). The six clusters together appear to have a significant lack of higher-mass 
stars in relation to the expected number from a Salpeter-like IMF and the total number
of stars present.  A larger statistical sample of small-N systems is required
to firmly establish this possibility.

\section{SUMMARY}

We can now construct a working model for the origin of the IMF based on the physical
processes known to occur in star formation and their effects determined through
numerical simulations (Fig.~\ref{expIMF}). This working model attributes the peak of the IMF and the characteristic
stellar mass to gravitational fragmentation and the thermal physics at the point of fragmentation.
Lower-mass stars and brown dwarfs are ascribed to fragmentation in dense regions and then
ejection to truncate the accretion rates while higher-mass stars are due to the continued
competitive accretion in the dense cores of forming stellar clusters. It is worth noting that
all three physical processes are primarily due to gravity and thus in combination 
provide the simplest
mechanism to produce the IMF.  

There is much work yet to be done in terms of including additional physics 
(magnetic fields, feedback)
into the numerical simulations that produce testable IMFs. It is also important
to develop  additional observational predictions from the theoretical models and to
use observed properties of star forming regions to determine necessary and sufficient conditions
for a full theory for the origin of the IMF. For example, competitive accretion predicts
that high-mass star formation is linked to the formation of a bound stellar cluster. 
This can be tested by observations: the existence
of significant numbers of high-mass stars in non-clustered regions or small-N clusters
would argue strongly against the accretion model.

\medskip
\centerline\textbf{\bf Acknowledgments}

\medskip

We thank the referee, Bruce Elmegreen, for valuable comments. 
HZ  thanks the DFG for travel support to PPV.

\vspace{0.1cm} 

\centerline\textbf{ REFERENCES}
\vspace{0.1cm} 
\parskip=0pt
{\small
\baselineskip=11pt

\refs Adams F.~C. and  Fatuzzo M.\ (1996) \apj, {\sl 464}, 256-271. 

\refs Arons J., and Max C.~E.\  (1975) \apj, {\sl 196}, L77-81.
 
\refs Ballesteros-Paredes J., V{\'a}zquez-Semadeni E. and Scalo J.\ (1999) 
\apj, {\sl 515}, 286-303.

\refs Ballesteros-Paredes J., Gazol A., Kim J., Klessen R.~S., Jappsen 
A.-K. and Tejero E.\ (2006) \apj, {\sl 637}, 384-391. 

\refs Bally J. and 
Zinnecker H.\ (2005) \aj, {\sl 129}, 2281-2293.

\refs Bastien P., Arcoragi 
J.-P., Benz W., Bonnell I. and Martel H.\ (1991) \apj, {\sl 378}, 255-265. 
 
\refs Basu S. and Jones  C.~E.\ (2004) \mnras, {\sl 347}, L47-51.

\refs Bate M.~R.\ (2005) \mnras, {\sl 363}, 
363-378.

\refs  Bate M.~R. and Bonnell I.~A.\ (1997) \mnras, {\sl 285}, 33-48. 
 
\refs  Bate M.~R. and  Bonnell I.~A.\ (2005) \mnras, {\sl 356}, 1201-1221.

\refs Bate M.~R. and  Burkert A.\ (1997) \mnras, {\sl 288}, 1060-1072. 

\refs Bate M.~R., Bonnell I.~A. and Price N.~M.\ (1995) \mnras, {\sl 277}, 362-376. 
 
\refs Bate M.~R., Bonnell  I.~A. and Bromm V.\ (2002a) \mnras, {\sl 332}, L65-68. 

\refs Bate M.~R., Bonnell I.~A. and Bromm V.\ (2002b) \mnras, {\sl 336}, 705-713.

\refs Bate M.~R., Bonnell I.~A. and Bromm V.\ (2003) \mnras, {\sl 339}, 577-599.
  

\refs Binney J. and 
Tremaine S.\ (1987), in {\sl Galactic Dynamics}, p 747. Princeton University, Princeton.

\refs  Bondi H. and Hoyle F.\ (1944) \mnras, {\sl 104}, 273-282.

\refs  Bonnell I.~A. and Bate M.~R.\ (2002) \mnras, {\sl 336}, 659-669.

\refs  Bonnell I.~A. and Bate M.~R.\ (2005) \mnras, {\sl 362}, 915-920.
 
\refs Bonnell I.~A. and Clarke C.~J.\ (1999) \mnras, {\sl 309}, 461-464.

\refs  Bonnell I.~A. and Davies M.~B.\ (1998) \mnras, {\sl 295}, 691-698.

\refs  Bonnell I.~A., Bate M.~R., Clarke C.~J. and Pringle J.~E.\ (1997) \mnras, {\sl 285}, 201-208.

\refs Bonnell I.~A., Bate M.~R. and Zinnecker H.\ (1998) \mnras, {\sl 298}, 93-102. 

\refs Bonnell I.~A., Bate M.~R., Clarke C.~J. and Pringle J.~E.\ (2001a) \mnras, {\sl 323}, 785-794. 

\refs  Bonnell I.~A., Clarke C.~J. Bate M.~R. and Pringle J.~E.\ (2001b) \mnras, {\sl 324}, 573-579.

\refs  Bonnell I.~A., Smith K.~W., Davies M.~B. and Horne K.\ (2001c) \mnras, {\sl 322}, 859-865.

\refs Bonnell I.~A., Bate M.~R. and Vine S.~G.\ (2003) \mnras, {\sl 343}, 413-418. 

\refs  Bonnell I.~A., Vine S.~G. and Bate M.~R.\ (2004) \mnras, {\sl 349}, 735-741.  

\refs  Bonnell I.~A., Clarke C.~J. and Bate M.~R.,  (2006a) \mnras, in press.

\refs  Bonnell I.~A., Dobbs C.~L., Robitaille T.~P. and Pringle J.~E.\ (2006b) \mnras, {\sl 365}, 37-45.

\refs Bonnor W.~B.\ (1956) \mnras, 
{\sl 116}, 351-359.

\refs  Boss A.~R.\ (1986) \apjs, {\sl 62}, 519-552.
 
\refs Boyd D.~F.~A. and 
Whitworth A.~P.\ (2005) \aap, {\sl 430}, 1059-1066. 

\refs  Burkert A. and 
Hartmann L.\ (2004) \apj, {\sl 616}, 288-300.
 
\refs Carpenter J.~M., 
Meyer M.~R., Dougados C., Strom S.~E. and Hillenbrand L.~A.\ (1997) \aj, 
{\sl 114}, 198-221.
 
\refs Chabrier G.\ (2003) \pasp, 
{\sl 115}, 763-795.

\refs  Clark P.~C. and Bonnell I.~A.\ (2005) \mnras, {\sl 361}, 2-16.   

\refs  Clark P.~C. and Bonnell I.~A.\ (2006) \mnras, submitted.

\refs Clarke C. J.\ (1998) in {\sl The Stellar Initial Mass Function: ASP 
Conf.~Ser.~142} (G. Gilmore and D. Howell eds), pp 189-199. ASP, San Francisco.

\refs Corbelli E., Palla F. and Zinnecker H.\ (2005) 
{\sl The Initial Mass Function 50 years 
later}. ASSL Vol {\sl 327}.~ Springer, Dordrecht.

\refs Dale J.~E., Bonnell I.~A., Clarke C.~J. and Bate M.~R.\ (2005) \mnras, {\sl 358}, 291-304. 
  
\refs de Wit W.~J., Testi 
L., Palla F. and Zinnecker H.\ (2005) \aap, {\sl 437}, 247-255. 
  
\refs Delgado-Donate E.~J., Clarke C.~J. and Bate M.~R.\ (2003) \mnras, {\sl 342}, 926-938.

\refs Dobbs C.~L., Bonnell I.~A. and Clark P.~C.\ (2005) \mnras, {\sl 360}, 2-8.

\refs Duch{\^e}ne G., 
Bouvier J., Bontemps S., Andr{\'e} P. and Motte F.\ (2004) \aap, {\sl 427}, 
651-665.

\refs  Duquennoy A. and 
Mayor M.\ (1991) \aap, {\sl 248}, 485-524.
  
\refs Durisen R.~H., 
Sterzik M.~F. and Pickett B.~K.\ (2001) \aap, {\sl 371}, 952-962.

\refs Ebert R.\ (1955) {\sl Zeitschrift fur 
Astrophysik, 37}, 217-232. 

\refs Edgar R. and Clarke C., (2004) \mnras, {\sl 349}, 
678-686.

\refs Edgar R.~G., 
Gawryszczak A. and Walch S.\ (2005) In {\it PPV
Poster Proceedings} \\ 
http://www.lpi.usra.edu/meetings/ppv2005/pdf/8005.pdf

\refs Elmegreen B.~G.\ (1993) 
\apj, {\sl 419}, L29-32. 

\refs  Elmegreen B.~G.\ (1997) 
\apj, {\sl 486}, 944-954.

\refs  Elmegreen B.~G.\ (1999) 
\apj, {\sl 515}, 323-336.
  
\refs  Elmegreen B.~G.\ (2004) 
\mnras, {\sl 354}, 367-374.
 
  
\refs Elmegreen B.~G. and Falgarone E.\ (1996) \apj, {\sl 471}, 816-821. 
 
\refs Elmegreen B.~G. and Krakowski A.\ (2001) \apj, {\sl 562}, 433-439. 

\refs Elmegreen B.~G. 
and Mathieu R.~D.\ (1983) \mnras, {\sl 203}, 305-315.

\refs Elmegreen B.~G. and Scalo J.\ (2004) \araa, {\sl 42}, 211-273. 

\refs Fleck R.~C.\ (1982) \mnras, {\sl 201}, 
551-559.

\refs  Figer D.~F.\ (2005) \nat, {\sl 434}, 192-194. 
 
\refs Garc{\'{\i}}a B. and Mermilliod J.~C.\ (2001) \aap, {\sl 368}, 122-136. 

\refs Goodwin S.~P. and 
Kroupa P.\ (2005) \aap, {\sl 439}, 565-569.
 
\refs  Heitsch F., Mac Low M.-M. and Klessen R.~S.\ (2001) \apj, {\sl 547}, 280-291. 
 
\refs Heyer M.~H. and Brunt C.~M.\ (2004) \apj, {\sl 615}, L45-48.

\refs  Hillenbrand L.~A. and Hartmann L.~W.\ (1998) \apj, {\sl 492}, 540-553.
 
\refs Hofmann K.-H., 
Seggewiss W. and Weigelt G.\ (1995) \aap, {\sl 300}, 403-414.

\refs Hubber D.~A., Goodwin 
S.~P. and Whitworth A.~P.\ (2006) \aap, in press.

\refs Jappsen A.-K., Klessen R.~S., Larson R.~B., Li Y. and Mac Low M.-M.\ (2005) \aap, {\sl 435}, 
611-623.

\refs  Johnstone D., 
Wilson C.~D., Moriarty-Schieven G., Joncas G., Smith G., Gregersen E. 
and Fich M.\ (2000) \apj, {\sl 545}, 327-339. 
 
\refs  Johnstone D., Di 
Francesco J., and Kirk H.\ (2004) \apj, {\sl 611}, L45-48. 
 
\refs Klessen R.~S.\ (2001) \apj, 
{\sl 556}, 837-846.
 
\refs  Klessen R.~S., 
Ballesteros-Paredes J., V{\'a}zquez-Semadeni E. and Dur{\'a}n-Rojas C.\ 
(2005) \apj, {\sl 620}, 786-794. 

\refs  Klessen R.~S. and 
Burkert A.\ (2000) \apjs, {\sl 128}, 287-319. 

\refs  Klessen R.~S. and 
Burkert A.\ (2001) \apj, {\sl 549}, 386-401. 

\refs  Klessen R.~S., 
Burkert A. and Bate M.~R.\ (1998) \apj, {\sl 501}, L205-208. 

\refs Kroupa P.\ (1995) \mnras, {\sl 277}, 
1491-1506. 

\refs  Kroupa P.\ (2001) \mnras, {\sl 322}, 
231-246.
 
\refs  Kroupa P.\ (2002) Science, {\sl 295}, 
82-91.

\refs  Krumholz M.~R.,  (2006) ApJ, submitted.

\refs Krumholz M.~R., 
McKee C.~F. and Klein R.~I.\ (2005a) \nat, {\sl 438}, 332-334.

\refs  Krumholz M.~R., 
McKee C.~F. and Klein R.~I.\ (2005b) \apj, {\sl 618}, L33-36.

\refs  Krumholz M.~R., 
Klein R.~I. and McKee, C.~F.\ (2005c) in {\sl Massive star birth: A crossroads of astrophysics: IAU Symposium 227} (R. Cesaroni \etal~ eds.)  pp 231-236. Cambridge University, Cambridge.

\refs  Krumholz M.~R., 
McKee C.~F., and Klein R.~I.\ (2006) \apj, {\sl 638}, 369-381.

\refs Lada C.~J.\ (2006) ArXiv 
Astrophysics e-prints, arXiv:astro-ph/0601375. 
 
\refs Lada C.~J. and Lada E.~A.\ (2003) \araa, {\sl 41}, 57-115. 

\refs Larson R.~B.\ (1978) \mnras, 
{\sl 184}, 69-85. 
  
\refs Larson R.~B.\ (1981) \mnras, 
{\sl 194}, 809-826.
   
\refs Larson R.~B.\ (1982) \mnras, 
{\sl 200}, 159-174. 
 
\refs Larson R.~B.\ (1985) \mnras, 
{\sl 214}, 379-398. 
  

\refs Larson R.~B.\ (1992) \mnras, 
{\sl 256}, 641-646. 

\refs Larson, R.~B.\ (2003) in {\sl Galactic Star Formation Across the Stellar Mass Spectrum} (J.M. De Buizer and N.S. van der Bliek eds.),
pp 65-80.  ASP, San Francisco.

\refs Larson R.~B.\ (2005) \mnras,  {\sl 359}, 211-222. 

\refs Layzer D.\ (1963) \apj, {\sl 137},  351-362.
 
\refs  Li Z.-Y. and Nakamura 
F.\ (2004) \apj, {\sl 609}, L83-86. 

\refs  Li Z.-Y. and Nakamura 
F.\ (2005) ArXiv Astrophysics e-prints, arXiv:astro-ph/0512278. 

\refs Littlefair S.~P. 
Naylor T., Jeffries R.~D., Devey C.~R. and Vine S.\ (2003) \mnras, {\sl 345}, 
1205-1211. 

\refs  Lizano S. and Shu F.~H.\ (1989) \apj, {\sl 342}, 834-854. 

\refs Low C. and Lynden-Bell D.\ (1976) \mnras, {\sl 176}, 367-390.

\refs  Mac Low M.-M. and 
Klessen R.~S.\ (2004) {\sl Reviews of Modern Physics,  76}, 125-194.
 
\refs  Mac Low M.-M., 
Klessen R.~S., Burkert A. and Smith M.~D.\ (1998) {\sl Physical Review 
Letters, 80}, 2754-2757.

\refs Malkov O. and 
Zinnecker H.\ (2001) \mnras, {\sl 321}, 149-154.

\refs  Mason B.~D., Gies 
D.~R., Hartkopf W.~I., Bagnuolo W.~G., Brummelaar T.~T. and McAlister 
H.~A.\ (1998) \aj, {\sl 115}, 821-847. 
 
 \refs Massi F., Testi L. and 
Vanzi L.\ (2005) ArXiv Astrophysics e-prints, arXiv:astro-ph/0511794. 

\refs Massey P. and 
Hunter D.~A.\ (1998) \apj, {\sl 493}, 180-194.
 
\refs  McDonald J.~M. and 
Clarke C.~J.\ (1993) \mnras, {\sl 262}, 800-804.

\refs  McKee C.~F. and Tan J.~C.\ (2003) \apj, {\sl 585}, 850-871. 
 
\refs Mestel L. and  Spitzer L.\ (1956) \mnras, {\sl 116}, 503-514.
 
\refs  Meyer M.~R., Adams 
F.~C., Hillenbrand L.~A., Carpenter J.~M. and Larson R.~B.\ 2000, 
in {\it Protostars and
Planets IV} (V. Mannings et al., eds.), pp. 121-150. Univ. of Arizona,
Tucson.

\refs Miller G.~E. and 
Scalo J.~M.\ (1979) \apjs, {\sl 41}, 513-547.

\refs Motte F., Andre P. and 
Neri R.\ (1998) \aap, {\sl 336}, 150-172.

\refs Motte F., Andr{\'e} P., Ward-Thompson D. and Bontemps S.\ (2001) \aap, {\sl 372}, L41-44.
 
\refs Nayakshin S. and  Sunyaev R.\ (2005) \mnras, {\sl 364}, L23-27.

\refs  Oey M.~S. and Clarke  C.~J.\ (2005) \apj, {\sl 620}, L43-47.
  
\refs  Ostriker E.~C., 
Gammie C.~F. and Stone J.~M.\ (1999) \apj, {\sl 513}, 259-274.
  
\refs Padoan P.\ (1995) \mnras, {\sl 277}, 
377-388.
  
\refs  Padoan P. and 
Nordlund {\AA}.\ (2002) \apj, {\sl 576}, 870-879.

\refs Padoan P. and 
Nordlund {\AA}.\ (2004) \apj, {\sl 617}, 559-564.
 
\refs Padoan P., Nordlund 
A. and Jones B.~J.~T.\ (1997) \mnras, {\sl 288}, 145-152.
 
\refs  Padoan P., Juvela M., 
Goodman A.~A. and Nordlund {\AA}.\ (2001) \apj, {\sl 553}, 227-234.

\refs Palla F., Randich S., 
Flaccomio E. and Pallavicini R.\ (2005) \apj, {\sl 626}, L49-52.

\refs Paumard T., et al.\ 
(2006) ArXiv Astrophysics e-prints, arXiv:astro-ph/0601268. 
  
\refs  Preibisch T., 
Balega Y., Hofmann K.-H., Weigelt G. and Zinnecker H.\ (1999) {\sl New 
Astronomy, 4}, 531-542. 
 
\refs  Price N.~M. 
and Podsiadlowski P.\ (1995) \mnras, {\sl 273}, 1041-1068.
 
\refs Rees M.~J.\ (1976) \mnras, {\sl 176}, 
483-486.
 
\refs  Reid I.~N., Gizis J.~E. 
and Hawley S.~L.\ (2002) \aj, {\sl 124}, 2721-2738.

\refs  Reipurth B. and 
Clarke C.\ (2001) \aj, {\sl 122}, 432-439.
 
\refs  Salpeter E.~E.\ (1955) \apj, {\sl 121}, 161-167.

\refs Sagar R. and 
Richtler,T.\ (1991) \aap, {\sl 250}, 324-339.

\refs  Scalo J.~M.\ (1986) Fundamentals 
of Cosmic Physics, {\sl 11}, 1-278.

\refs Scalo J.~M.\ (1998) in {\sl The Stellar Initial Mass Function: ASP 
Conf.~Ser.~142} (G. Gilmore and D. Howell eds), pp 201-236. ASP, San Francisco.

\refs  Scalo J.~M.\  (2005)  in {\sl The Initial Mass Function 50 years 
later} (E.~Corbelli, F.~Palla, and H. Zinnecker eds.) pp 23-39. ASSL Vol {\sl 327}.~ Springer, 
Dordrecht.

\refs Schmeja S. and 
Klessen R.~S.\ (2004) \aap, {\sl 419}, 405-417.
 
\refs Shu, F.~H. Adams F.~C. 
and Lizano S.\ (1987) \araa, {\sl 25}, 23-81.
 
\refs  Shu F.~H., Li Z.-Y. and 
Allen A.\ (2004) \apj, {\sl 601}, 930-951.

\refs Silk J.\ (1982) \apj, {\sl 256}, 514-522.

\refs Silk J.\ (1995) \apj, {\sl 438}, L41-44.
 
\refs Silk J. and 
Takahashi T.\ (1979) \apj, {\sl 229}, 242-256.

\refs Spaans M. and Silk J.\ (2000) \apj, {\sl 538}, 115-120.  
 
\refs Stanke T., McCaughrean 
M.~J. and Zinnecker H.\ (2000) \aap, {\sl 355}, 639-650.

\refs Sterzik M.~F. and 
Durisen R.~H.\ (1998) \aap, {\sl 339}, 95-112.
 
\refs Sterzik M.~F. and 
Durisen R.~H.\ (2003) \aap, {\sl 400}, 1031-1042.

\refs Sterzik M.~F., 
Durisen R.~H. and Zinnecker H.\ (2003) \aap, {\sl 411}, 91-97.

 \refs  Stone J.~M., Ostriker 
E.~C. and Gammie C.~F.\ (1998) \apj, {\sl 508}, L99-102.

\refs  Stolte A., Grebel 
E.~K., Brandner W. and Figer D.~F.\ (2002) \aap, {\sl 394}, 459-478.

\refs Stolte A., Brandner 
W., Grebel E.~K., Lenzen R. and Lagrange A.-M.\ (2005) \apj, {\sl 628}, L113-117. 

\refs Stolte A., \etal~(2006) in preparation.
 
\refs  Testi L. and 
Sargent A.~I.\ (1998) \apj, {\sl 508}, L9-94. 
 
\refs  Testi L., Palla F. and 
Natta A.\ (1999) \aap, {\sl 342}, 515-523.

\refs  Tilley D.~A. and 
Pudritz R.~E.\ (2005) ArXiv Astrophysics e-prints, arXiv:astro-ph/0508562. 
 
\refs  Tohline J.~E.\ (1982) 
Fundamentals of Cosmic Physics, {\sl 8}, 1-81.

\refs V{\'a}zquez-Semadeni E.\  (1994) \apj, {\sl 423}, 681-692.

\refs V{\'a}zquez-Semadeni E., Ostriker E.~C., Passot T., Gammie C.~F. and Stone 
J.~M.\ (2000) in {\sl Protostars and Planets IV} (V. Mannings \etal  eds.), pp 3-28 University of Arizona Press, Tucson.

\refs V{\'a}zquez-Semadeni E., Kim J. and Ballesteros-Paredes J.\ (2005) \apj, 
{\sl 630}, L49-52.

\refs  Weidner C. and Kroupa P.\ (2004) \mnras, {\sl 348}, 187-191.

\refs  Weidner C. and Kroupa P.\ (2006) \mnras, {\sl 365}, 1333-1347.

%
\refs  Wolfire M.G. and Cassinelli J.P., (1987) ApJ, {\sl 319}, 850-867.

\refs Yoshii Y. and Saio 
H.\ (1985) \apj, {\sl 295,} 521-536.

\refs  Yorke H. and Sonnhalter C., (2002) ApJ, {\sl 569}, 846-862.

\refs Zinnecker H.\ (1982) {\sl New 
York Academy Sciences Annals,  395}, 226-235.

\refs Zinnecker H.\ (1984) \mnras, 
{\sl 210}, 43-56.
 
\refs Zinnecker H.\  (2003) in 
       {\sl A Massive Star Odyssey: From Main Sequence to Supernova: IAU Symposium 212}
                (K.v.d. Hucht, A. Herrero, and C. Esteban eds.), pp. 80-90. ASP, San Francisco.
     
\refs Zinnecker H.\  (2005) in {\sl The Young Local Universe, XXXIXth
Rencontres de Moriond} (A. Chalabaev,
Th. Montmerle, and J. Tran Thanh Van eds), p. 17-22. Th\'e Gi\'oi, Vietnam.

 \refs Zinnecker H., 
McCaughrean M.~J. and Wilking B.~A.\ (1993) in {\sl Protostars and Planets III}
(E. H. Levy and J. I. Lunine, eds.), pp. 429-495. Univ. of Arizona, Tucson.

\end{document}